\documentclass[aip,jcp,superscriptaddress]{revtex4-1}
\usepackage[dvips]{graphics}
\usepackage{graphicx}
\usepackage{amsfonts}
\usepackage{amssymb}
\usepackage{amsmath}
\usepackage{subfigure}

\begin{document}

\title{Control of the symmetry breaking in double-well potentials by the
resonant nonlinearity management}
\author{H. E. Nistazakis}
\affiliation{Department of Physics, University of Athens, Panepistimiopolis, Zografos, Athens 15784, Greece}
\author{B. A. Malomed}
\affiliation{Department of Physical Electronics,\ School of Electrical Engineering,
Faculty of Engineering, Tel Aviv University, Tel Aviv 69978, Israel}
\author{P. G. Kevrekidis}
\affiliation{Department of Mathematics and Statistics, University of Massachusetts,
Amherst, MA 01003-4515}
\author{D. J. Frantzeskakis}
\affiliation{Department of Physics, University of Athens, Panepistimiopolis, Zografos, Athens 15784, Greece}

\begin{abstract}
We introduce a one-dimensional model of Bose-Einstein condensates (BECs),
combining the double-well potential, which is a well-known setting for the
onset of spontaneous-symmetry-breaking (SSB) effects, and time-periodic
modulation of the nonlinearity, which may be implemented by means of the
Feshbach-resonance-management (FRM) technique. Both cases of the
nonlinearity which is repulsive or attractive on the average are considered.
In the former case, the main effect produced by the application of the FRM
is spontaneous self-trapping of the condensate in either of the two
potential wells in parameter regimes where it 
would remain untrapped in the absence of the
management. In the weakly nonlinear regime, 
the frequency of intrinsic oscillations in the FRM-induced
trapped state is very close to half the FRM frequency, suggesting that the
effect is accounted for by a parametric resonance. In the case of the
attractive nonlinearity, the FRM-induced effect is the opposite, i.e.,
enforced detrapping of a state which is self-trapped in its unmanaged form.
In the latter case, the frequency of oscillations of the untrapped mode is
close to a quarter of the driving frequency, suggesting that a higher-order
parametric resonance may account for this effect.
\end{abstract}

\pacs{03.75.Lm; 42.65.Wi; 03.75.Lm}
\maketitle

{\bf Symmetric double-well potentials (DWPs) play an important role
in many physical situations dominated by self-focusing and
defocusing nonlinearities. It is well known that, in quantum
mechanics, the DWP gives rise to alternating symmetric and
antisymmetric bound states. However, the nonlinearity breaks the
symmetry, giving rise to asymmetric bound states. This effect of the
spontaneous symmetry breaking (SSB) has drawn a great deal of
attention in theoretical and experimental studies of nonlinear
physical media, especially in the context of Bose-Einstein
condensation (BEC). In this work, we aim to extend the analysis of
the SSB by introducing a BEC model which combines the DWP and the
time-periodic modulation of the strength of the nonlinearity. The
latter is another important tool used for the control of the
nonlinear dynamics in BEC, by means of the physical effect known as
the Feshbach resonance (controlled by a time-periodic external
field, in that case). We consider the interplay of the DWP and
time-modulated Feshbach resonance ({\it Feshbach-resonance
management}, FRM) for both signs of the nonlinearity, repulsive and
attractive. In these two cases, new dynamical effects are predicted
by means of a systematic numerical analysis. For the repulsive (on
the average) nonlinearity, the application of the FRM induces {\it
self-trapping} of the condensate in one of the two symmetric
potential wells, in the case when the free condensate would be
untrapped. In such an FRM-induced trapped state, 
for low density cases, 
the frequency of
intrinsic oscillations is found to be very close to half the
underlying FRM driving frequency, which suggests that the trapping
is induced by a parametric resonance. In the opposite case of the
attractive (on the average) nonlinearity, the effect of the FRM is
opposite too, namely it is {\it detrapping} of states which were
self-trapped in the absence of the FRM, with the frequency of
oscillations of the detrapped condensate close to a quarter of the
driving frequency 
for low density cases too. 
A higher-order parametric resonance may plausibly
account for the latter effect.}

\section{Introduction}

The past two decades have witnessed remarkable progress in experimental and
theoretical studies of Bose-Einstein condensates (BECs) \cite{book1,book2}.
Nonlinear matter-wave structures have been one of the significant subjects
considered in this context \cite{ourbook,revnonlin,spin1}. From the theoretical
viewpoint, the emergence of such structures can be understood in the
framework of the well-established mean-field approximation, based on the
Gross-Pitaevskii equation. In particular, for attractive (repulsive)
interatomic interactions, this equation predicts the existence of bright
(dark) matter-wave solitons, which have been observed in a series of
experiments, see Refs.~\cite{expb1,expb2,expb3} and \cite%
{dark1,dark2,dark3,dark4,hamburg,hambcol,kip,technion,draft6}, respectively.
Bright gap solitons are also possible in BEC with repulsive interactions
loaded into an optical lattice \cite{gap}.

An important element in the studies of BEC is the variety of external
potentials that may be used to confine the ultracold atomic gases. The most
typical forms of such trapping potentials are the harmonic-oscillator traps
and optical lattices, i.e., periodic potentials created by the interference
of counter-propagating laser beams. A similar situation is relevant also in
the context of nonlinear optics, where the basic model, namely the
nonlinear Schr\"{o}dinger equation, may incorporate harmonic or periodic
potentials describing graded-index waveguides and periodic waveguiding
arrays, respectively \cite{kivshar}. A combination of harmonic and periodic
traps may be used to create a double-well potential (DWP), which has drawn a
great deal of attention. BECs confined in DWPs were studied experimentally in
Refs.~\cite{markus1,markus11}, 
and many theoretical works addressed this setting 
\cite{smerzi,kiv2,mahmud,bam,Bergeman_2mode,infeld,todd,theo,carr}. 
A fundamental effect studied, in diverse forms, in these works, is the spontaneous
symmetry breaking (SSB) of the population of atoms in the two symmetric
wells. A related work was done in nonlinear optics, where twin-core
self-guided laser beams in Kerr media \cite{HaeltermannPRL02}, optically
induced dual-core waveguiding structures in photorefractive crystals 
\cite{zhigang}, and trapped light beams in an annular core of an optical fiber
\cite{Longhi} among others, also lead to manifestations of phenomenology
associated with DWPs.

In addition to the studies of the SSB in the static DWP settings,
the analysis was extended to diverse situations with the DWP subject
to ``management", i.e., its parameters, such the depth of the well
or height of the barrier between them, were made periodically
varying functions of time \cite{DWP-management}. In fact, in all
those works the analysis was based on a two-mode approximation,
which replaces the underlying Gross-Pitaevskii equation (GPE) by a
system of two coupled ordinary differential equations (ODEs).

Apart from the variety in the shapes of the potentials, another important
tool for manipulations of BEC is provided by magnetically \cite%
{Koehler,feshbachNa} or optically \cite{ofr} induced Feshbach resonance,
which makes it possible to control the effective nonlinearity in the
condensate. The latter possibility has given rise to many theoretical and
experimental studies. A well-known example is the formation of bright
matter-wave solitons and soliton trains in $^{7}$Li \cite{expb1,expb2} and $%
^{85}$Rb \cite{expb3} condensates, by switching the interatomic
interactions from repulsive to attractive. Many theoretical works
studied the BEC dynamics under temporal and/or spatial modulation of
the nonlinearity. In particular, the application of such a
``Feshbach resonance management\textquotedblright\ (FRM) technique
in the temporal domain can be
used to stabilize attractive higher-dimensional BEC against collapse \cite%
{Isaac,FRM1}, and to create robust matter-wave breathers in the
effectively one-dimensional (1D) condensate \cite{FRM2}. On the
other hand, the so-called ``collisionally inhomogeneous
condensates\textquotedblright , controlled by the spatially
modulated nonlinearity, have been predicted to support a variety of
new effects, such as the adiabatic compression of matter-waves
\cite{our1,fka}, Bloch oscillations of matter-wave solitons
\cite{our1}, atomic soliton emission and atom lasers \cite{vpg12},
dynamical trapping of matter-wave solitons \cite{HS,our2},
enhancement of transmissivity of matter-waves through barriers
\cite{our2,fka2}, stable condensates exhibiting both attractive and
repulsive interatomic interactions \cite{HS,chin}, the
delocalization transition~\cite{LocDeloc}, SSB in a nonlinear
double-well pseudopotential \cite{Dong}, the competition between
incommensurable linear and nonlinear lattices \cite{HSnew}, and
others.

In this work we combine the two above-mentioned settings, namely the DWP and
a time-modulated nonlinearity, in the 1D geometry, with the objective to
control the SSB in the double-well setting by means of the FRM. Our model is
based on the corresponding GPE, see Eqs. (\ref{1D}) and (\ref{U1D}) below.
We consider the nonlinearity which may either be repulsive or attractive on
the average. In the case of the repulsive interactions, we find that the use
of the FRM results in the onset of spontaneous self-trapping of the BEC in
one of the two potential wells, while without the application of the FRM
atoms oscillate between the wells, i.e., the condensate is untrapped. We
also find that, when the self-trapping occurs, the trapped BEC undergoes
small-amplitude intrinsic oscillations at a frequency equal to half the
nonlinearity-modulation frequency, which indicates the role of a parametric
resonance in this case, 
when the densities are sufficiently low. 
In the case of the attractive nonlinearity, the
FRM-induced\ effect turns out to be the opposite: while the unmanaged
condensate is spontaneously trapped in one of the two potential wells, the
application of the FRM leads to its detrapping (dynamical \textit{symmetry
restoration}), with the density oscillations between the wells at a
frequency which is almost exactly a quarter of the driving frequency 
for low density cases. 
Thus,
for either nonlinearity (repulsive or attractive), our results show that the
FRM can control the BEC\ dynamics in DWPs, by inducing trapping/detrapping
and intrinsic oscillations, which do not take place in the absence of the FRM. 
Our numerical results, obtained in the framework of the GPE model, are 
also corroborated by a semi-analytical approximation. In particular, we 
extend the analytical approximation presented in Refs.~\cite{todd,theo,zhigang} 
to the FRM-driven setting under consideration: we adopt a Galerkin-type 
expansion to describe the evolution of the wavefunctions at 
each well of the DWP. This leads to a non-autonomous system of ODEs, which 
are solved numerically. In all cases, the ODE result is compared to 
the one obtained by the GPE model and, in most cases, they are found to be 
in fairly good agreement.

It is relevant to mention that a similar physical model was recently
introduced in Ref. \cite{xie}, where it was postulated that, under the
action of the FRM driving, the DWP was described by a system of linearly
coupled ODEs corresponding to the above-mentioned two-mode approximation.
However, dynamical regimes studied in Ref. \cite{xie} were fairly
different from those considered here. The analysis reported in that work
demonstrated an FRM-induced transition from untrapped oscillations to a
trapped state, but this was observed in the {\it high-frequency} regime, 
and was explained by means of an averaging method. In the present work, the
transition to the trapping is clearly accounted for by a parametric
resonance at {\it moderate} values of the FRM frequency $\omega $.
%, while a
%trapping transition at large $\omega $ was not observed in the simulations
of the underlying GPE. 
On the other hand, the ODE system used in Ref.~\cite{xie} 
cannot explain the parametric resonance (clearly, a more 
sophisticated approximation is needed to capture the resonant
mechanism). Another difference is that we observe the FRM-induced
trapping and \emph{detrapping} effects (the latter was not reported
in Ref. \cite{xie}) in the regime of relatively weak FRM, when the 
time-dependent nonlinearity coefficient, $g(t)$, does not change its sign. 
In Ref. \cite{xie}, the opposite regime of the strong management was 
considered. 
%In fact, our simulations of the GPE produce only chaotic dynamics 
%in that case.

The rest of the paper is structured as follows. In Section II we introduce
the model, and present the results for the repulsive and attractive
nonlinearity in Sections III and IV, respectively. Section V concludes the
paper.

\section{The model and its consideration}

\subsection{The model}

In the mean-field approximation, the combination of the DWP and time-varying
nonlinearity is described by the GPE for the macroscopic wavefunction $\psi
\left( x,t\right) $, written in the scaled form:
\begin{equation}
i\psi _{t}=-\left( 1/2\right) \psi _{xx}+U\left( x\right) \psi +g(t)|\psi
|^{2}\psi -\mu \psi.
\label{1D}
\end{equation}
Here $\mu $ is the chemical potential, while the DWP is taken in the form of
\cite{theo,gener1}
\begin{equation}
U(x)=\left( 1/2\right) \Omega ^{2}x^{2}+V_{0}\mathrm{sech}^{2}\left(
x/w\right),  \label{U1D}
\end{equation}
where $\Omega $ is the normalized strength of the harmonic trap, while $V_{0}
$ and $w$ denote, respectively, the amplitude and width of the barrier used
(in combination with the parabolic trap) to create the DWP. Finally, the FRM
modulation function $g(t)$ is taken in the customary form \cite{FRM1},
\begin{equation}
g(t)=g_{0}+g_{1}\sin \left( \omega t\right) .  \label{g(t)}
\end{equation}

The same model may be interpreted in terms of nonlinear optics, with $t$
being the propagation distance, $\mu $ the propagation constant, and $g(t)$
representing a periodic modulation of the Kerr coefficient \cite{Isaac}. In
that case, potential (\ref{U1D}) defines two waveguiding channels in a
planar medium.

We aim to study localized FRM-driven states in the 1D setting described by
Eqs.~(\ref{1D}) and (\ref{U1D}).
%Our analysis starts with either
In particular, we numerically solve the above model with initial conditions
of the form of either an
antisymmetric (with respect to the two wells) initial
%configuration
state and $g_{0}>0$ (the self-repulsive nonlinearity, on the average), or a symmetric
%initial configuration
state and $g_{0}<0$ (i.e., the nonlinearity which is
self-attractive, on the average). This choice stems from the fact that both
configurations are well known to feature the SSB, in the absence of the
``management" \cite{smerzi,kiv2,mahmud,bam,Bergeman_2mode,infeld,todd,theo,carr}.

Numerical results, which are reported below, were obtained by means
of the split-step Fourier method. We use the following values of
parameters which adequately represent the generic situation: the
effective parabolic trap strength $\Omega $ assumes value $0.1$, the
chemical potential, $\mu $, ranges from $0.05$ to $0.5$, the
barrier's width is $w=0.5$, and its height is $V_{0}=1$. As concerns
the nonlinearity-modulation function, $g(t)$, the ``dc
value\textquotedblright\ $g_{0}$ is taken as $\pm 1$ for the
repulsive or attractive interatomic interactions, respectively,
while the modulation strength, $g_{1}$, varies in the interval of
$(0,1.5)$.
Finally, the driving frequency takes values in the interval of $0<\omega <1$.
In the context of BEC, such parameters may correspond to the $^{23}$Na or
a $^{7}$Li condensate, for the repulsive ($g_{0}=+1$) and attractive ($%
g_{0}=-1$) nonlinearity, respectively, confined in the trap with
longitudinal and transverse confining frequencies $2\pi \times 10$ Hz and $%
2\pi \times 100$ Hz. In these two cases, the numbers of $^{23}$Na and $^{7}$%
Li atoms are, respectively, $N\simeq 500$ and $1000$. The corresponding peak
values of the normalized peak densities (at the DWP minima) are $\left(
|\psi |^{2}\right) _{\max }=0.1$ and $0.17$ in the repulsive case, or $0.13$
for the case of the attraction, for the respective chemical potential is $%
0.22$, $0.3$ or $0.05$.

\subsection{Semi-analytical approach}

The spectrum of the Schr{\"o}dinger equation Eq.~(\ref{1D}) for $g(t)=0$, consists
of a ground state, $\psi_{0}(x)$, and excited states, $\psi_{l}(x)$ ($l \ge 1$). In
the weakly nonlinear regime of the full 
problem, using a Galerkin-type approach,
we expand $\psi(x,t)$ as \cite{theo},
\begin{eqnarray}
\psi(x,t)=c_0(t) \psi_0(x) + c_1(t) \psi_1(x)+\cdots,
\label{beq3}
\end{eqnarray}
and truncate the expansion, keeping solely the first two modes;
here $c_{0,1}(t)$ are unknown time-dependent complex prefactors. Once again,
it is worth noticing that
such an approximation (involving the truncation of higher-order modes and
the spatio-temporal factorization of the wavefunction),
is expected to be quite useful for a weakly nonlinear analysis, i.e.,
for a sufficiently small $L^2$ norm (or, physically speaking, number of
atoms) of the solution.

Substituting Eq.~(\ref{beq3}) into Eq.~(\ref{1D}), and projecting
the result onto the corresponding eigenmodes, we obtain the
following ordinary differential equations (ODEs), for the
coefficients of the projection onto $\psi_0$ and $\psi_1$,
respectively:
\begin{eqnarray}
i \dot{c}_0 &=& (\omega_0-\mu) c_0 - (g_0+g_1(t)) A_0 |c_0|^2 c_0 - (g_0+g_1(t)) B (2 |c_1|^2 c_0 + c_1^2 \bar{c}_0),
\label{beq4}
\\
i \dot{c}_1 &=& (\omega_1-\mu) c_1 - (g_0+g_1(t)) A_1 |c_1|^2 c_1 - (g_0+g_1(t)) B (2 |c_0|^2 c_1 + c_0^2 \bar{c}_1).
\label{beq5}
\end{eqnarray}
In Eqs. (\ref{beq4})-(\ref{beq5}), dots denote time derivatives, the
overbar stands for the complex conjugate, $\omega_{0,1}$ are
eigenvalues corresponding to the eigenstates $\psi_{0,1}$, while
$A_0=\int \psi_0^4 dx$, $A_1=\int \psi_1^4 dx$, and $B=\int \psi_0^2
\psi_1^2 dx$ are constants. Additional overlapping integrals, {\it
viz}., $\int \psi_0 \psi_1^3 dx$ and $\int \psi_1 \psi_0^3 dx$,
which formally appear in the course of the derivation of the ODE
system, vanish due to the opposite parities of the real wavefunctions
$\psi_0$ and $\psi_1$, in the framework of the underlying
linear Schr{\"o}dinger problem with the symmetric potential.

We now use amplitude-phase variables, $c_j=\rho_j e^{i \phi_j}$,
$j=0,1$ (the amplitudes $\rho_j$ and phases $\phi_j$ are assumed to be real), to derive
from the ODEs (\ref{beq4})-(\ref{beq5}) a set of four equations.
Introducing function $\varphi \equiv \phi_1-\phi_0$, we obtain the
equations for $\rho_0$ and $\phi_0$:
\begin{eqnarray}
\dot{\rho}_0 &=& (g_0+g_1(t)) B \rho_1^2 \rho_0 \sin(2 \varphi),
\label{beq6}
\\
\dot{\phi}_0 &=& (\mu-\omega_0) - (g_0+g_1(t)) A_0 \rho_0^2 - 2 (g_0+g_1(t)) B \rho_1^2 - (g_0+g_1(t)) B \rho_1^2 \cos(2 \varphi),
\label{beq7}
\end{eqnarray}
while the equations for $\rho_1, \phi_1$ are
\begin{eqnarray}
\dot{\rho}_1 &=& -(g_0+g_1(t)) B \rho_0^2 \rho_1 \sin(2 \varphi),
\label{beq6a}
\\
\dot{\phi}_1 &=& (\mu-\omega_1) - (g_0+g_1(t)) A_0 \rho_1^2 - 2 (g_0+g_1(t)) B \rho_0^2 - (g_0+g_1(t)) B \rho_0^2 \cos(2 \varphi),
\label{beq7a}
\end{eqnarray}
The conservation of the norm $N=\int |\psi|^2 dx$
implies $\rho_0^2+ \rho_1^2=N$. Finally, subtracting Eqs.~(\ref{beq7})
%for $\dot{\phi}_0$, and the corresponding one
and (\ref{beq7a}),
%for $\dot{\phi}_1$,
we obtain:
\begin{eqnarray}
\dot{\varphi} =&-&\Delta \omega + (g_0 + g_1(t)) (A_0 \rho_0^2 -  A_1 \rho_1^2) - (g_0 + g_1(t)) B (2 + \cos(2 \varphi)) (\rho_0^2-\rho_1^2).
\label{beq9}
\end{eqnarray}
where $\Delta \omega \equiv \omega_1-\omega_0$. Equations (\ref{beq6}), (\ref{beq6a}) and (\ref{beq9}) constitute a 
non-autonomous dynamical system, which
we solve numerically. This way, from $(\rho_0,\phi_0)$ and $(\rho_1,\phi_1)$ we can respectively find $c_0(t)$ and $c_1(t)$ and, thus,
obtain the wavefunction profile as per the approximation of Eq.~(\ref{beq3}). For a given set of parameter values, the outcome of this
calculation will be compared to the outcome of the direct numerical integration of Eq.~(\ref{1D}).

Below we consider, at first, the case of the repulsive nonlinearity,
and separately analyze the situations corresponding to
``small\textquotedblright\ and ``large\textquotedblright\ initial
peak densities, with the corresponding values $\left( |\psi
|^{2}\right) _{\max }=0.1$ and $0.17$.

\section{The repulsive nonlinearity}

\subsection{``Small'' initial values of the peak density}

\begin{figure}[tbh]
\includegraphics[width=8cm]{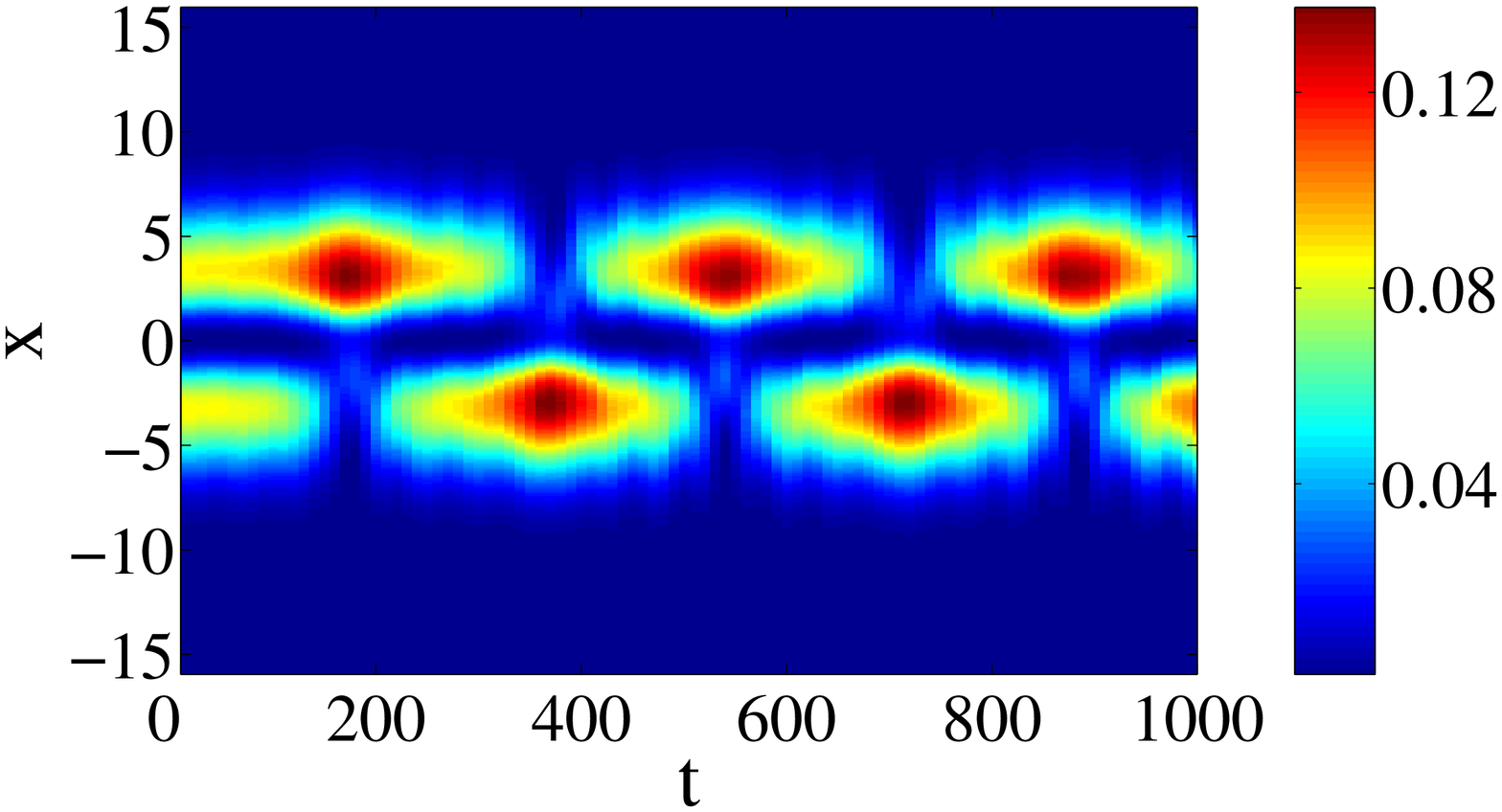} %
\includegraphics[width=6.5cm]{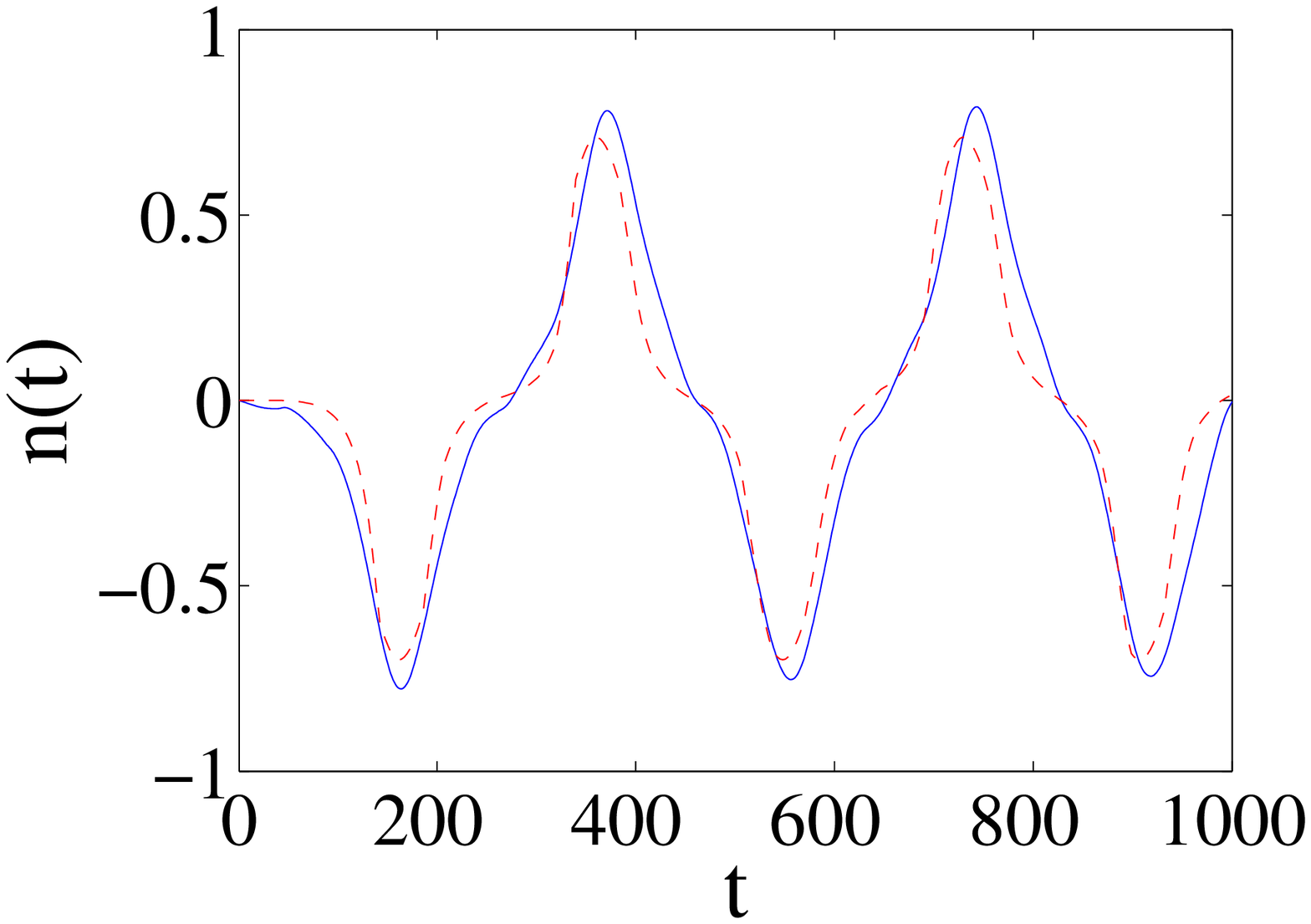}%{traj_01_g_0.eps}
\caption{(Color online) Left panel: The spatiotemporal evolution of
the density in the double-well potential. Right panel: the evolution
of the relative difference between the numbers of atoms in the left
and right wells, $n(t)$. The solid line represents the results from
the integration of the GPE, while the dashed one the results
obtained from the ODEs. In this case, the nonlinearity is repulsive,
$g_{0}=1$, with $g_{1}=0$ and initial value $\left( |\protect\psi
|^{2}\right) _{\max }=0.1$, while parameters of the potential are
$\Omega =0.1$, $V_{0}=1$ and $w=0.5$.} 
\label{fig1}
\end{figure}

In the case of the self-repulsive nonlinearity, we start by
considering the case of the ``unmanaged" condensate ($g_{1}=0$),
with parameters taken as in Ref. {\cite{theo}}, \textit{viz}.,
$g_{0}=1$, $\Omega =0.1$, $V_{0}=1$, $w=0.5$, and initial value
$\left( |\psi |^{2}\right) _{\max }=0.1$. Moreover, the parameters
involved in Eqs. (\ref{beq6}), (\ref{beq6a}) and (\ref{beq9}) are
found to be $A_0 = 0.09078$, $A_1=0.09502$, $B=0.08964$,
$\omega_0=0.13282$ and $\omega_1=0.15571$. In the left panel of Fig.
(\ref{fig1}) we show a spatiotemporal contour plot of the evolution
of the density, while the right panel of the same figure displays
the time evolution of the relative difference in the atomic population between the left
and the right wells, defined as 
\begin{equation}
n(t)=\frac{\int_{-\infty }^{0}|\psi |^{2}dx-\int_{0}^{+\infty }|\psi |^{2}dx%
}{\int_{-\infty }^{+\infty }|\psi |^{2}dx}.  \label{n_t}
\end{equation}%
Note that in the right panel of Fig. (\ref{fig1}) we show $n(t)$ as obtained from the 
GPE (solid line) and the ODEs, Eqs. (\ref{beq6}), (\ref{beq6a}) and (\ref{beq9}) (dashed
line); the agreement between the two is very good. A variable similar to $n(t)$ was 
also used in the recent work \cite{xie}, which analyzed the present setting in the 
framework of the ODEs of a two-mode reduction (see e.g. Ref.~\cite{smerzi}). 
The free oscillations of $n(t)$ in the untrapped state feature a well-defined frequency, which can be
estimated from the right panel of Fig. \ref{fig1} as $\omega _{\mathrm{osc}%
}\approx 0.021$. Moreover, from the value of $n(t)$ one can
straightforwardly estimate the fraction of the total number of atoms which
oscillates between the two wells and which is greater as the value of $n(t)$
approaches the $\pm 1$. In the case of Fig. {\ref{fig1}} the maximum value
of $\left\vert n(t)\right\vert $ is $0.84$.

The next step is to vary the FRM frequency, $\omega $, at different fixed
values of the FRM strength, $g_{1}$. Starting with $g_{1}=0.2$, we varied $%
\omega $ from $0$ to $1$ with step $1\times 10^{-4}$. The result was that,
in the intervals of $0\leq \omega <0.0793$ and $\omega \geq 0.1677$, the
oscillations of $n(t)$ remain periodic, like in Fig. {\ref{fig1}}, with
almost the same frequency as at $g_{1}=0$, i.e. $\omega _{\mathrm{osc}%
}\approx 0.021$, and with the zero average value of $n(t)$, i.e., the
managed system remains in the\emph{\ }untrapped state.

\begin{figure}[tbh]
\includegraphics[width=8cm]{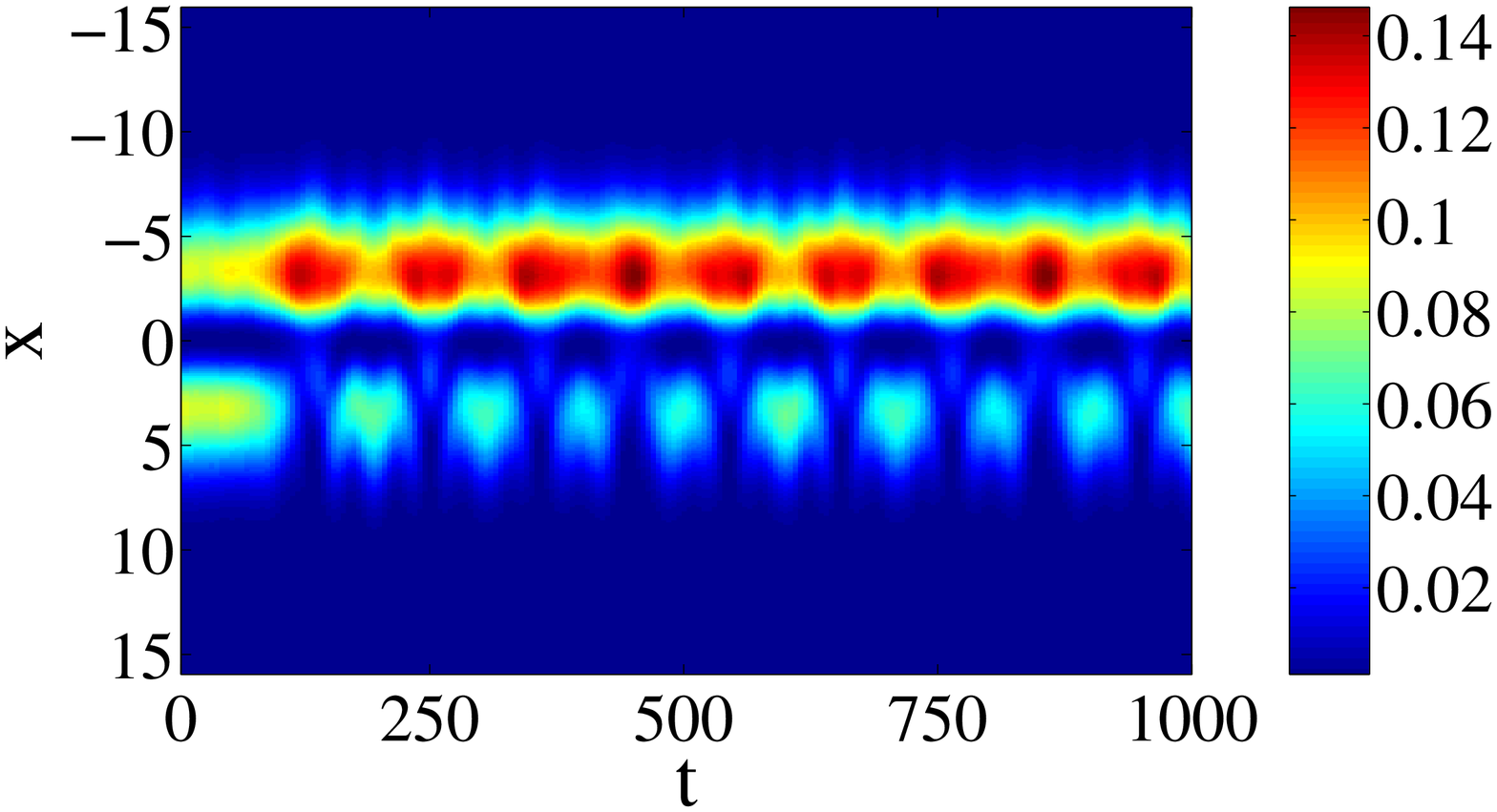}%{fig2a_new.eps} %
\includegraphics[width=8cm]{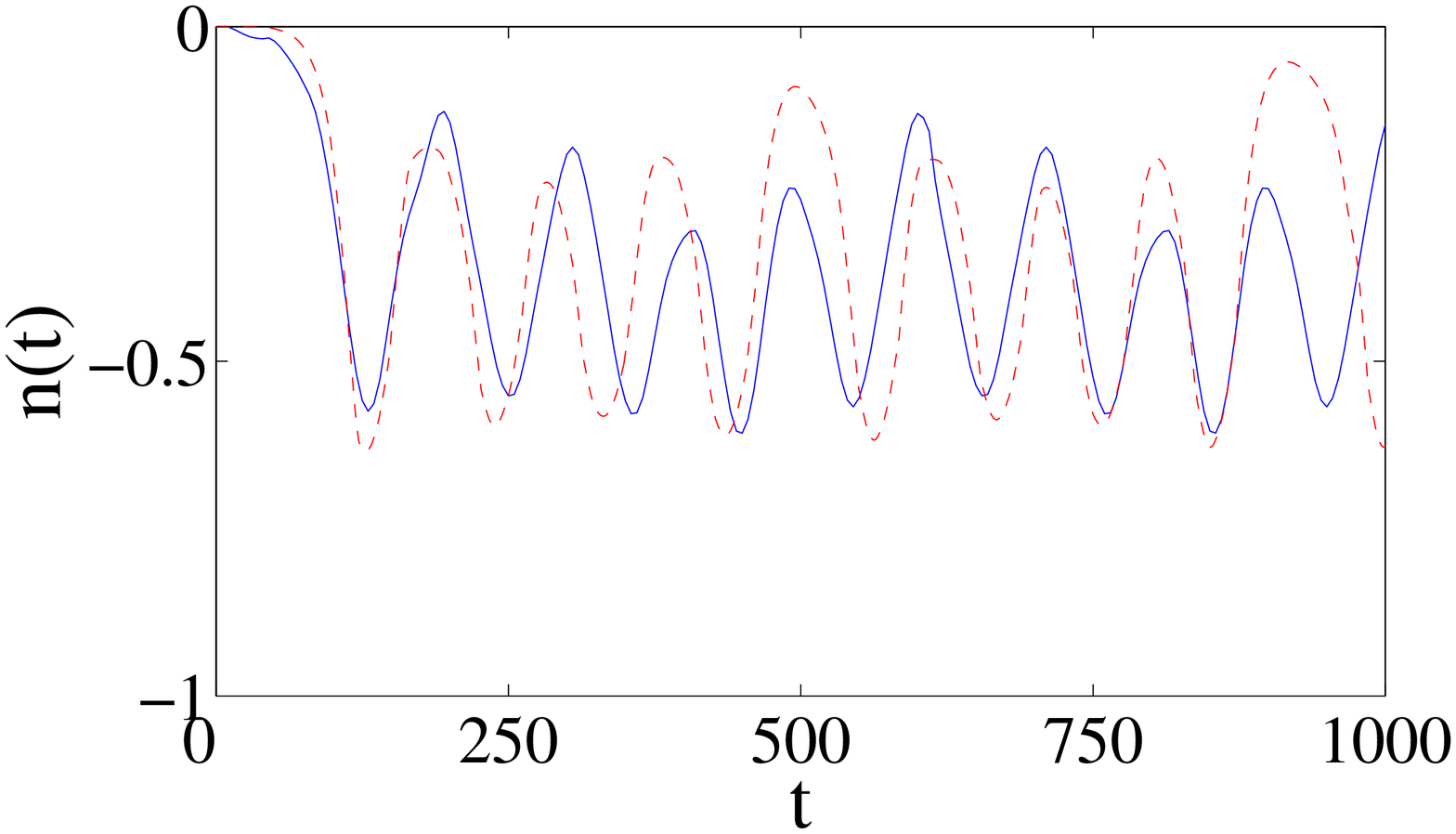}%{tr_rep_g1_02_om_011.eps}%{fig2b.eps}
\caption{(Color online) The same as in Fig. \protect\ref{fig1}, but in the
presence of FRM with $g_{1}=0.2$, $\protect\omega =0.11$, and initial value $%
\left( |\protect\psi |^{2}\right) _{\max }=0.1$. In the right panel,
the solid and dashed lines represent the results obtained from the
integration of the GPE and ODEs, respectively. Other parameters are
as in Fig. \protect\ref{fig1}.} \label{fig2}
\end{figure}

\begin{figure}[tbh]
\includegraphics[width=6.5cm]{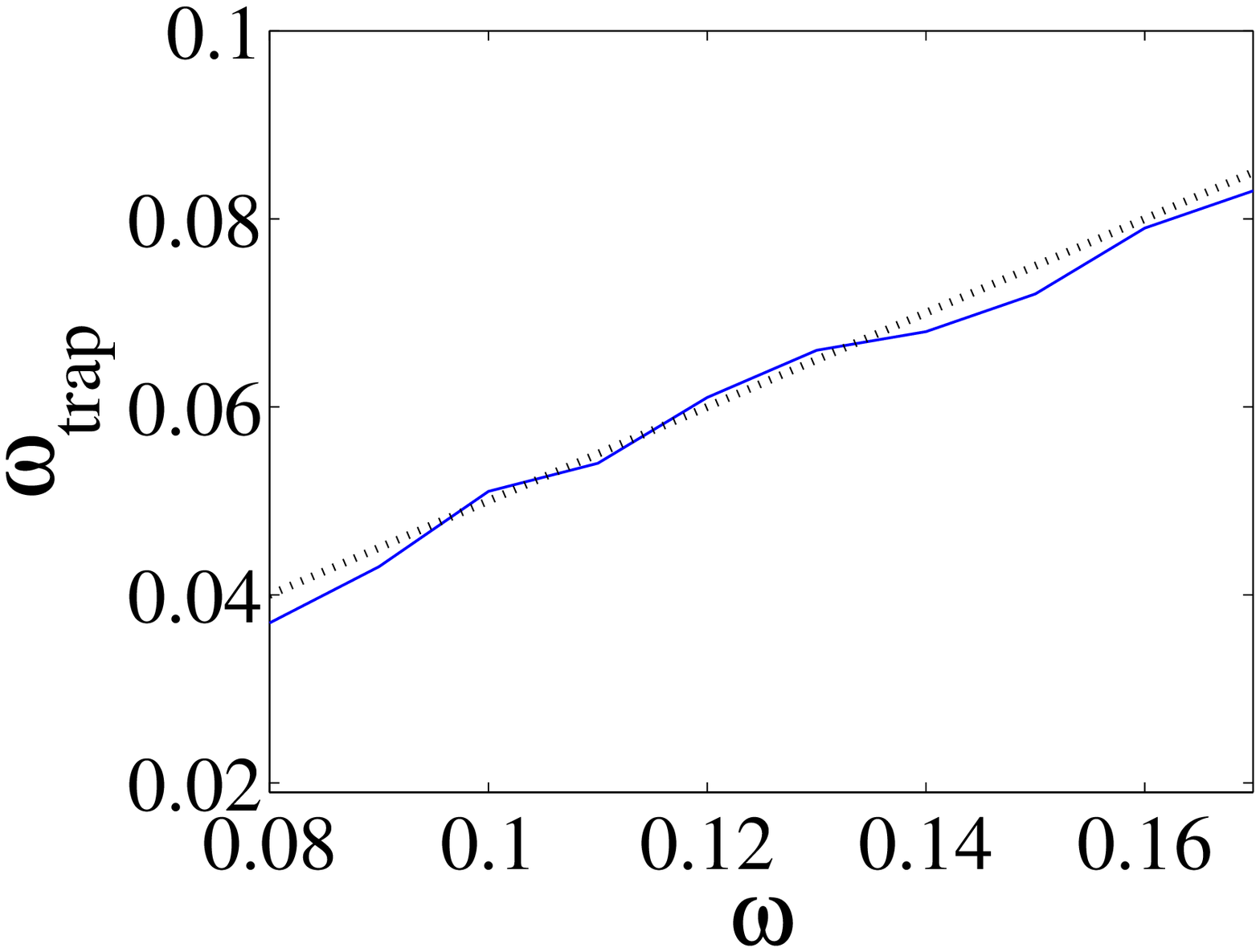}%{traj02_m022_new.eps} %
\includegraphics[width=8cm]{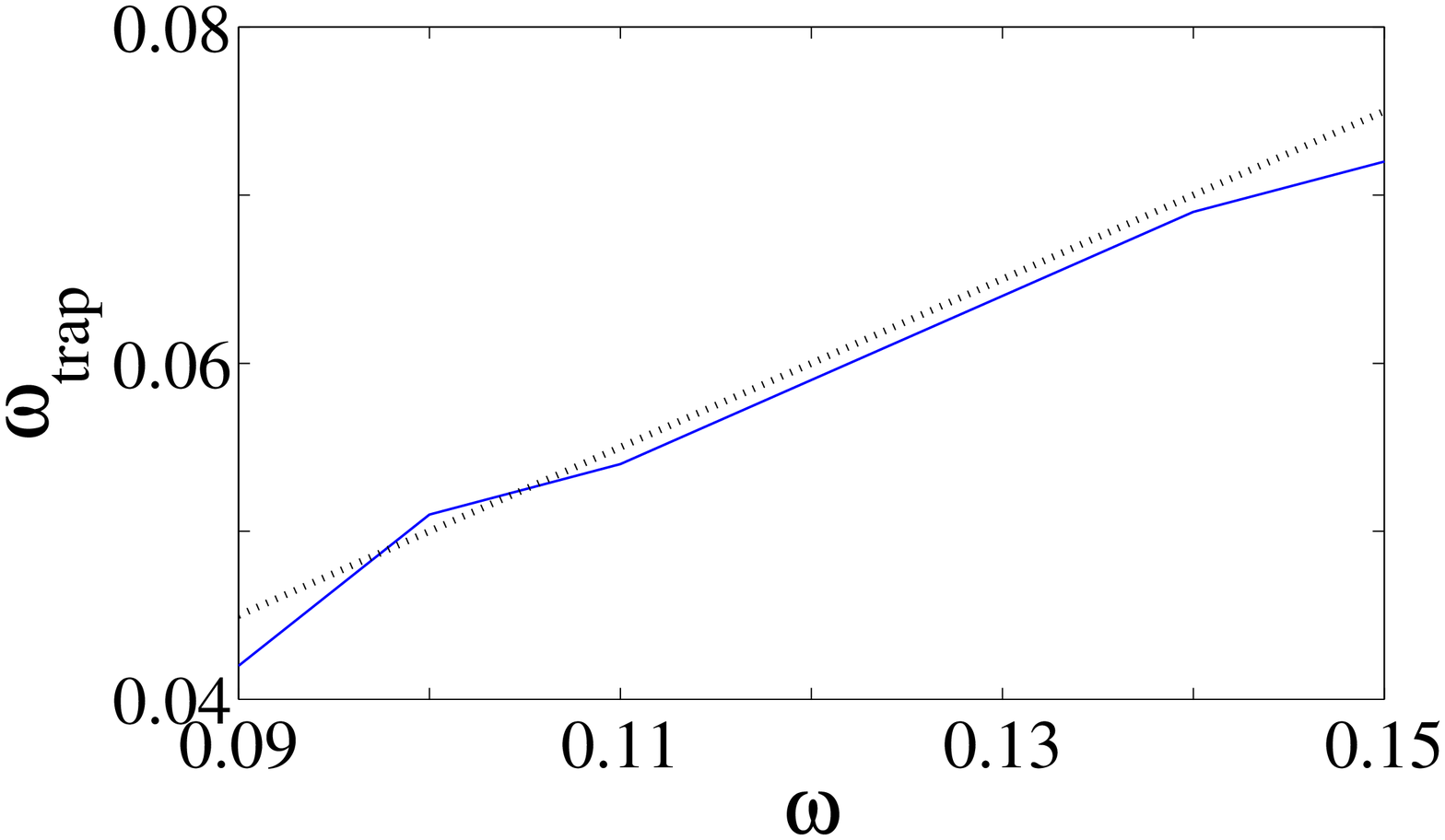}%{traj04_m022_new.eps}
\caption{(Color online) The relation between the FRM \ frequency, $\protect%
\omega $, and frequency $\protect\omega _{\mathrm{trap}}$ of the
oscillations of the condensate spontaneously trapped in one well under the
action of the FRM, in the case of the repulsive nonlinearity. 
Solid, and dotted lines correspond, respectively, to the result obtained 
from the GPE, and the estimation of $\omega _{\mathrm{trap}}= \omega /2$. 
Left panel: $%
g_{1}=0.2$ and
%$0.0793\leq \protect\omega <0.1677$
$0.08\leq \protect\omega <0.1677$. Right panel: $g_{1}=0.4$ and
%$0.0863\leq \protect\omega <0.1581$
$0.09\leq \protect\omega <0.15$. Other parameters are as in Figs. \protect
\ref{fig1} and \protect\ref{fig2}.}
\label{traj2}
\end{figure}

However, the example displayed in Fig. {\ref{fig2}} for $\omega =0.11$
demonstrates that the situation is completely different for
\begin{equation}
0.0793\leq \omega <0.1677,~\mathrm{i.e.,}~4\omega _{\mathrm{osc}}\lesssim
\omega \lesssim 8\omega _{\mathrm{osc}}~,  \label{trapped}
\end{equation}
when the the system gets \emph{trapped} in either of the two wells,
thus exhibiting an FRM-induced SSB. Actually, for all the values of
$\omega $ from interval (\ref{trapped}), $n(t)$ does not change its
sign, while the largest values of $\left\vert n(t)\right\vert $ are
above $0.5$, hence the observed trapping is quite strong. In the example shown
in Fig. {\ref{fig2},} the observed frequency of the trapped
oscillations is very close to half the
FRM frequency: $\omega _{\mathrm{trap}}=0.054\approx \omega /2$ (note that $%
\omega _{\mathrm{trap}}$ is very different from the above-mentioned
frequency of the oscillations in the untrapped state, $\omega _{\mathrm{osc}%
}\approx 0.021$). Examining the trapped states for other values of $\omega $
from interval (\ref{trapped}), we have concluded that relation $\omega _{%
\mathrm{trap}}\approx \omega /2$ remains valid as long as the FRM-induced
trapping holds, as seen in the left panel of Fig. \ref{traj2}, which
displays the dependence between the values of $\omega $ and $\omega _{%
\mathrm{osc}}$ in the domain of $4\omega _{\mathrm{osc}}\leq \omega <8\omega
_{\mathrm{osc}}$. This relation clearly suggests that the SSB trapping
induced by the FRM is a result of a parametric resonance \cite{resonance}.
To the best of our knowledge, this is the first example of the resonant SSB
in the DWP model, induced by the time-periodic drive. It is relevant to
mention that the parametric resonance also plays a dominant role in effects
induced by the management in the form of the time-periodic modulation of the
strength of the single-well trapping potential \cite{BBB}.

Next, increasing the value of the FRM amplitude to $g_{1}=0.4$, the results
are as follows: for $0\leq \omega <0.0863$ and $\omega \geq 0.1581$, the
condensate remains untrapped, with $n(t)$ featuring the oscillation
frequency which is virtually identical to that observed in the absence of
the FRM (i.e. $\omega _{\mathrm{osc}}\approx 0.021$). On the other hand, in
the interval of $0.0863\leq \omega <0.1581$, which is almost identical to $%
4\omega _{\mathrm{osc}}\leq \omega <8\omega _{\mathrm{osc}}$, cf. its
approximate counterpart in Eq. (\ref{trapped}), the oscillations of $n(t)$
demonstrate the SSB leading to a state trapped in one potential well with
maximum values of $\left\vert n(t)\right\vert $ greater than $0.5$. In this
case too, we have examined the relation between the FRM frequency, $\omega $%
, and the frequency of the trapped oscillations, $\omega _{\mathrm{trap}}$.
The obtained results corroborate that the above-mentioned
parametric-resonance relation, $\omega _{\mathrm{trap}}\approx \omega /2$,
remains valid, as seen in the right panel of Fig. \ref{traj2}.

\begin{figure}[tbh]
\includegraphics[width=5cm]{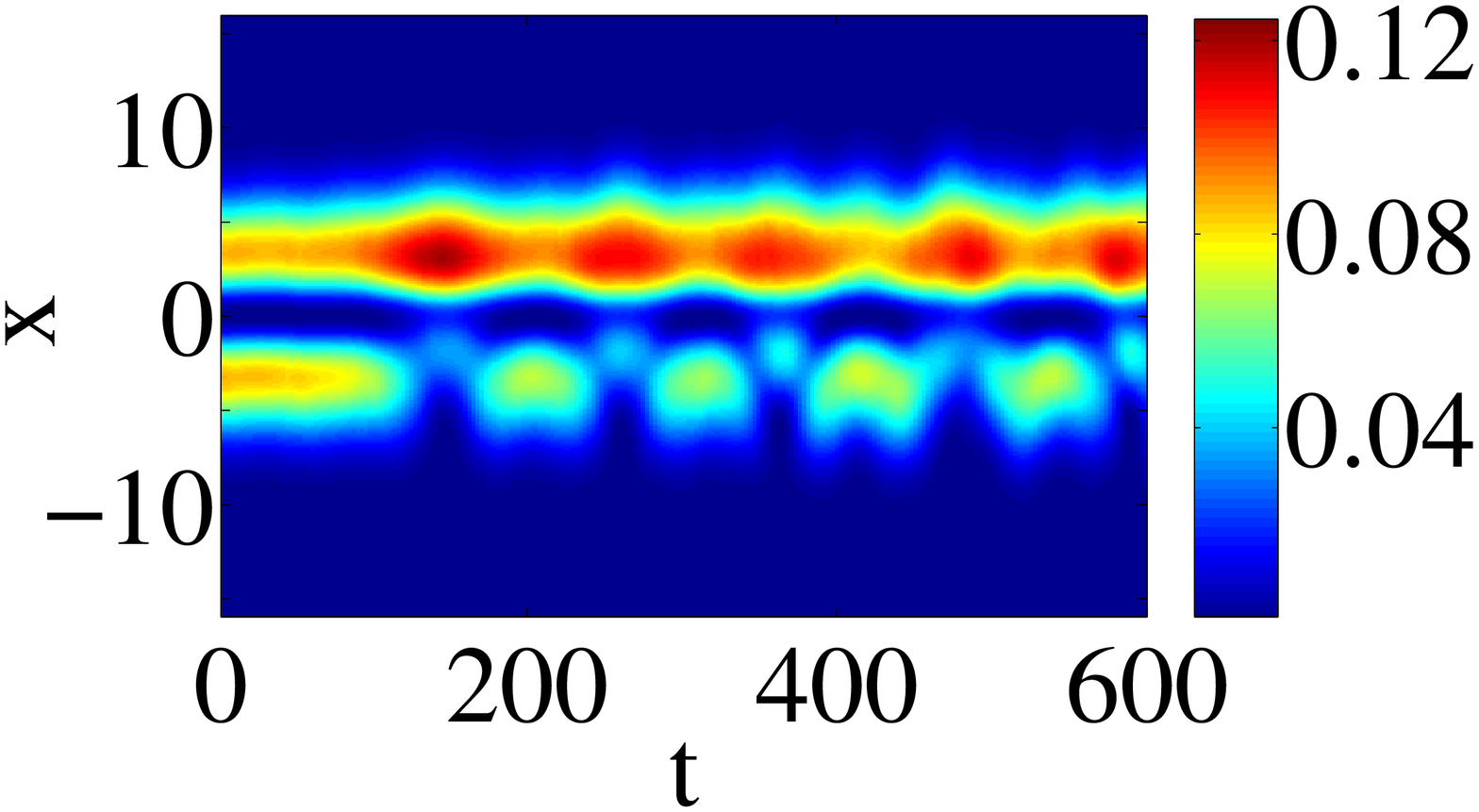} %
\includegraphics[width=5cm]{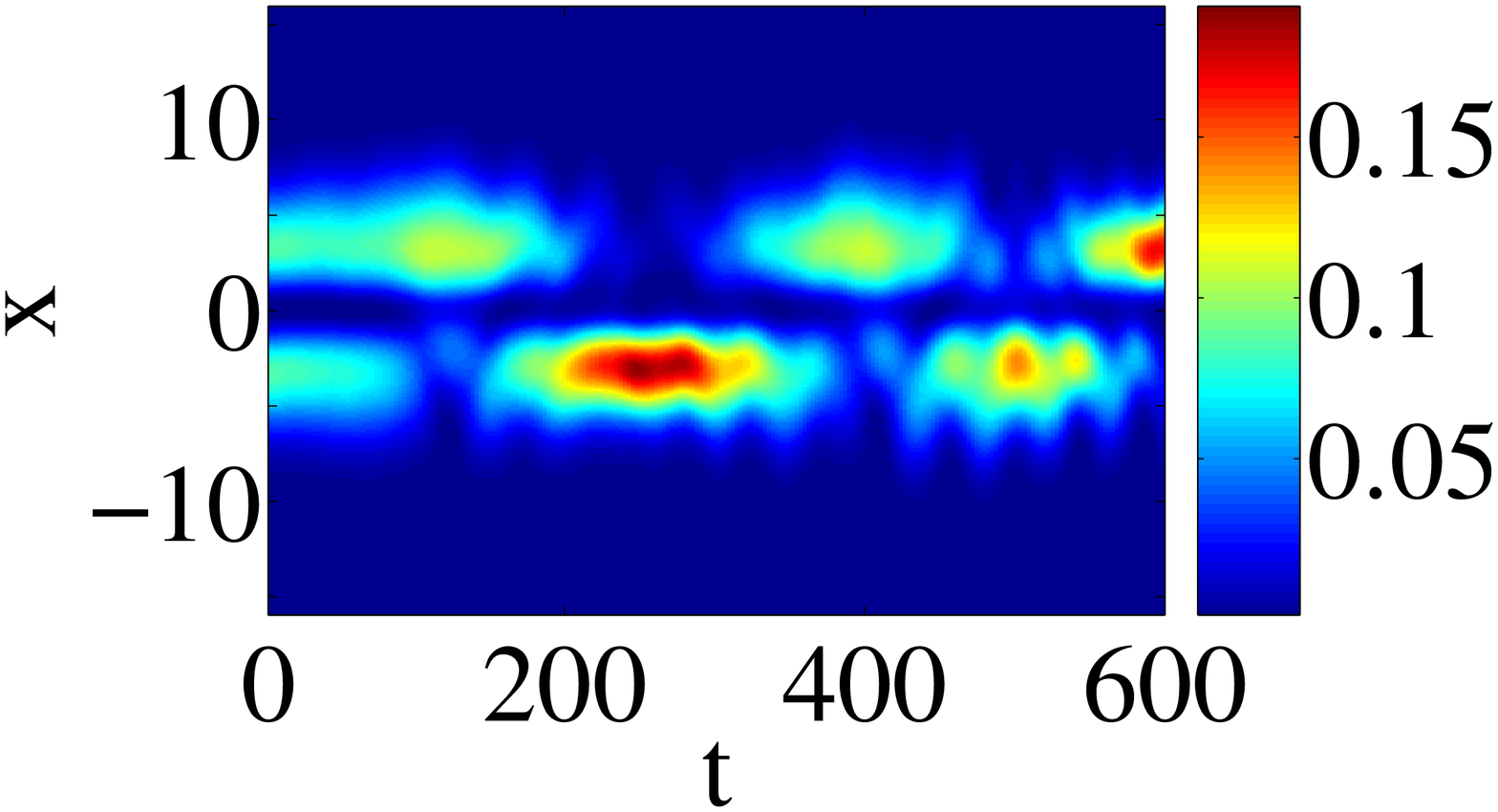} %
\includegraphics[width=5cm]{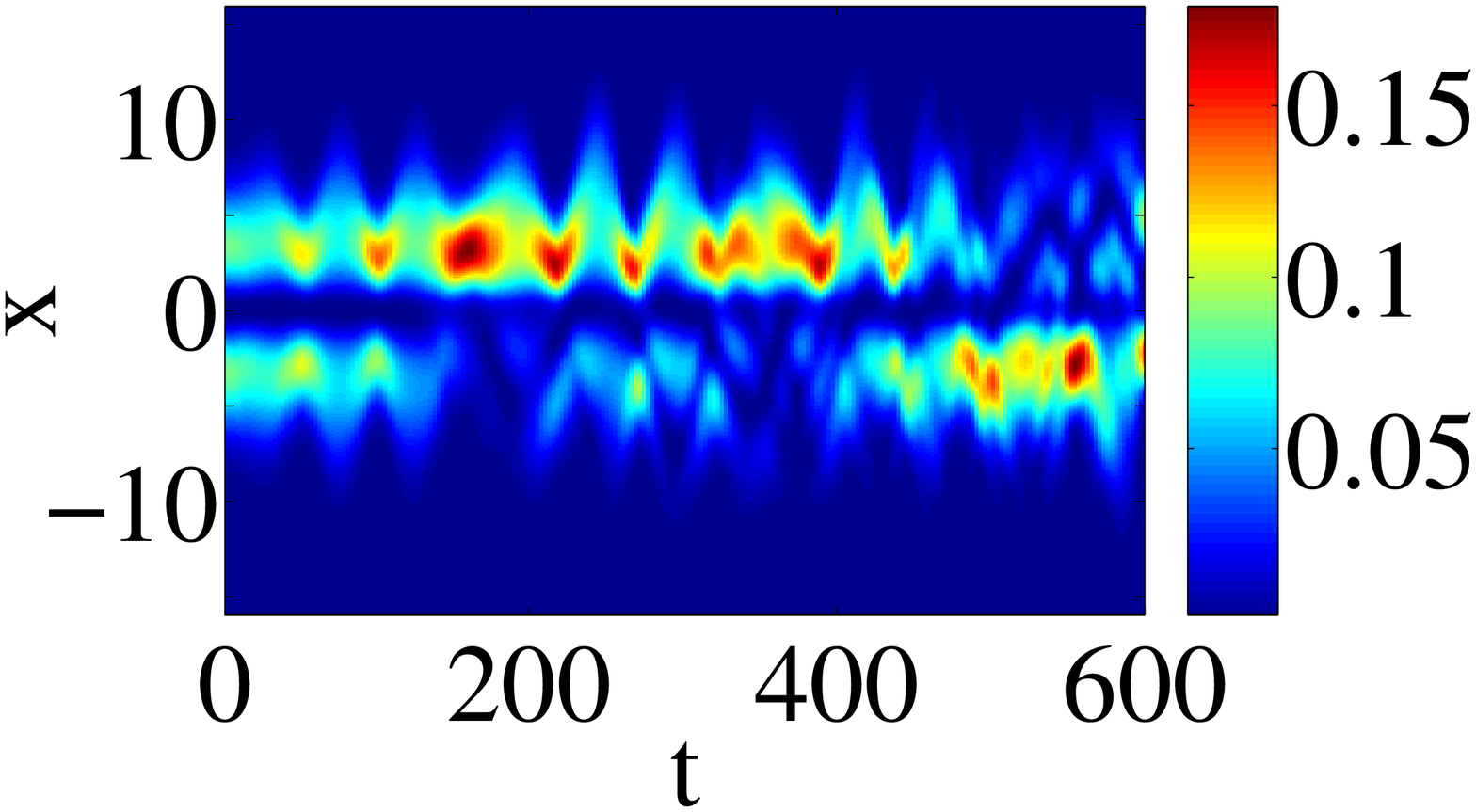} %
\includegraphics[width=5cm]{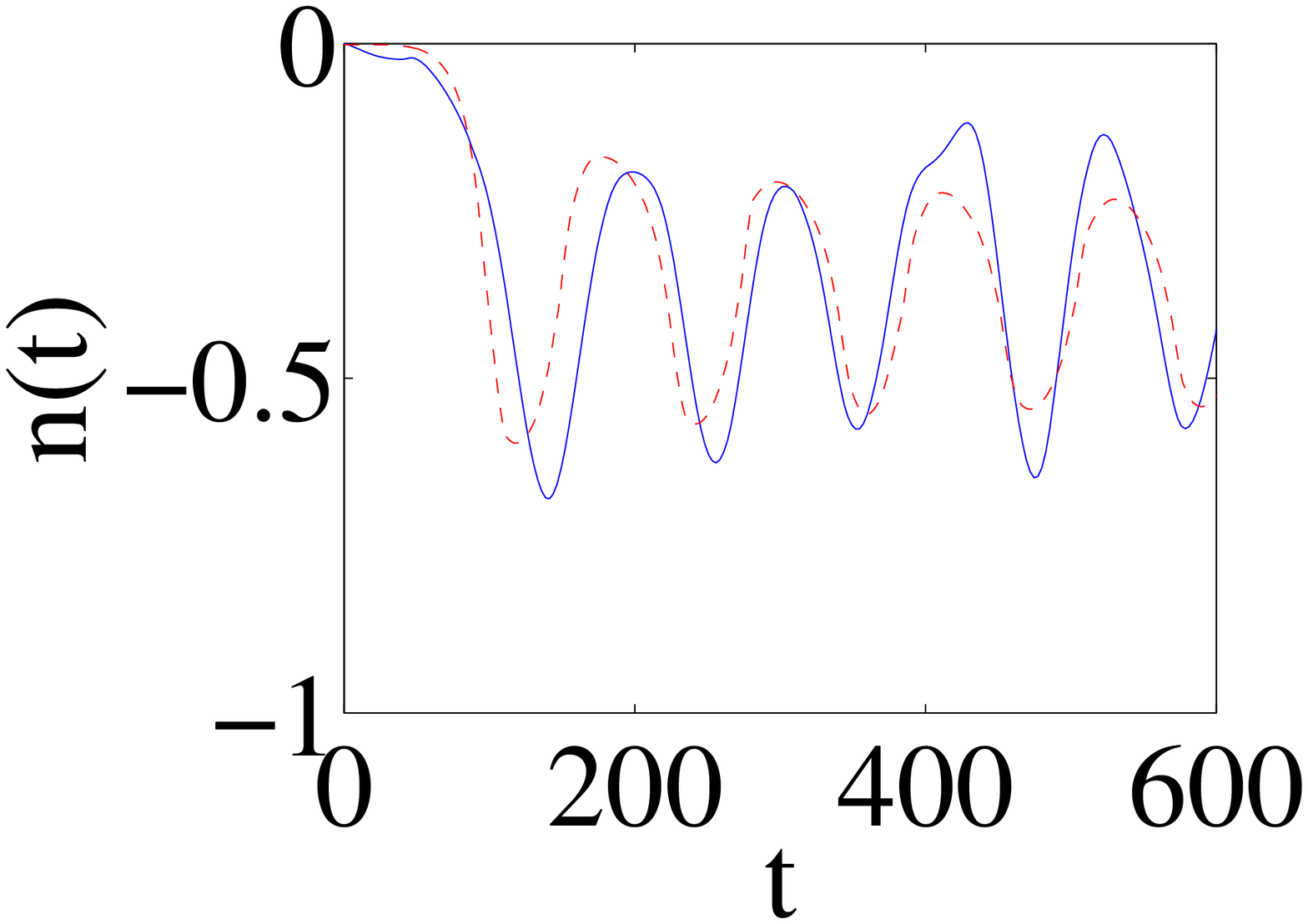} %
\includegraphics[width=5cm]{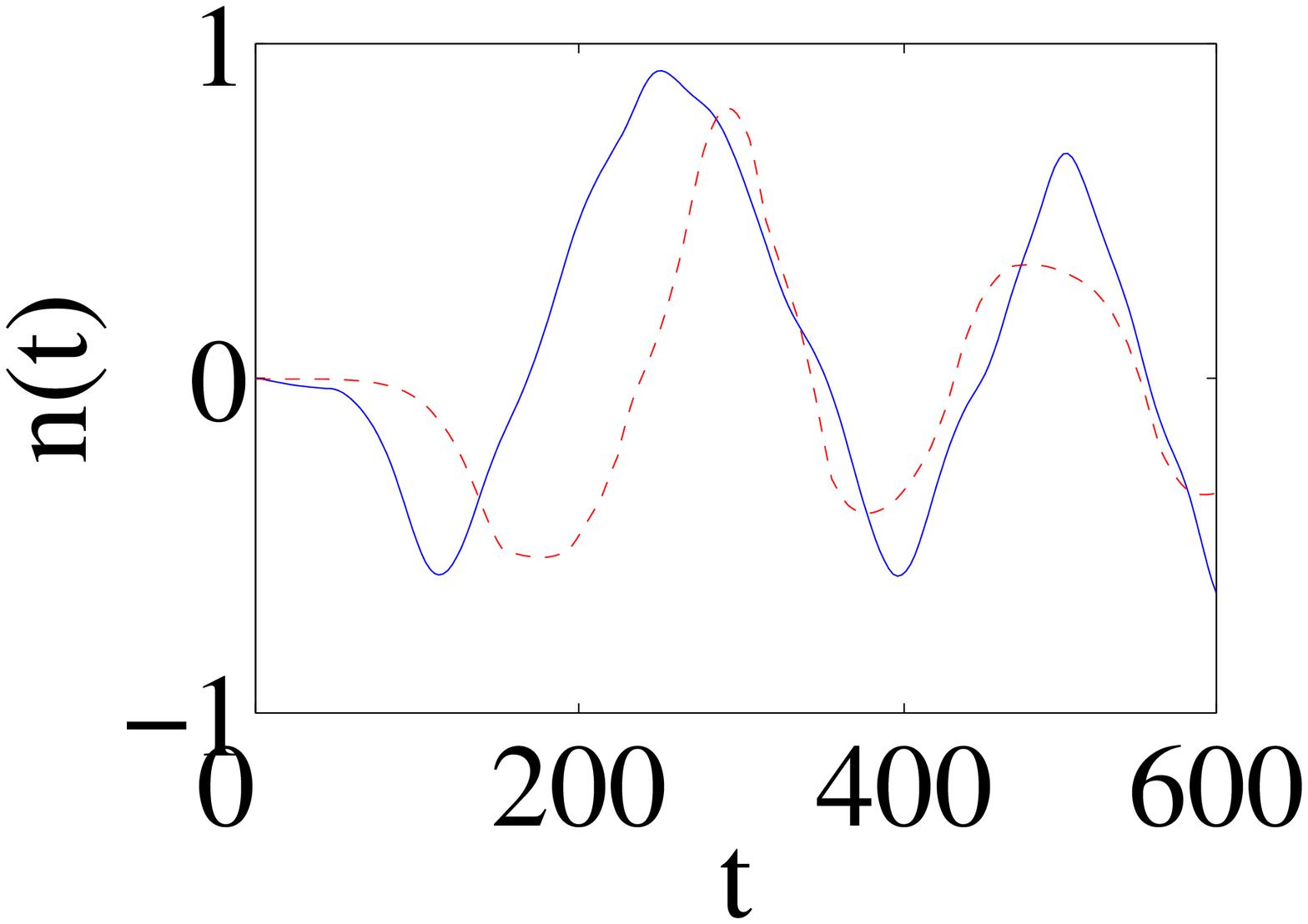} %
\includegraphics[width=5cm]{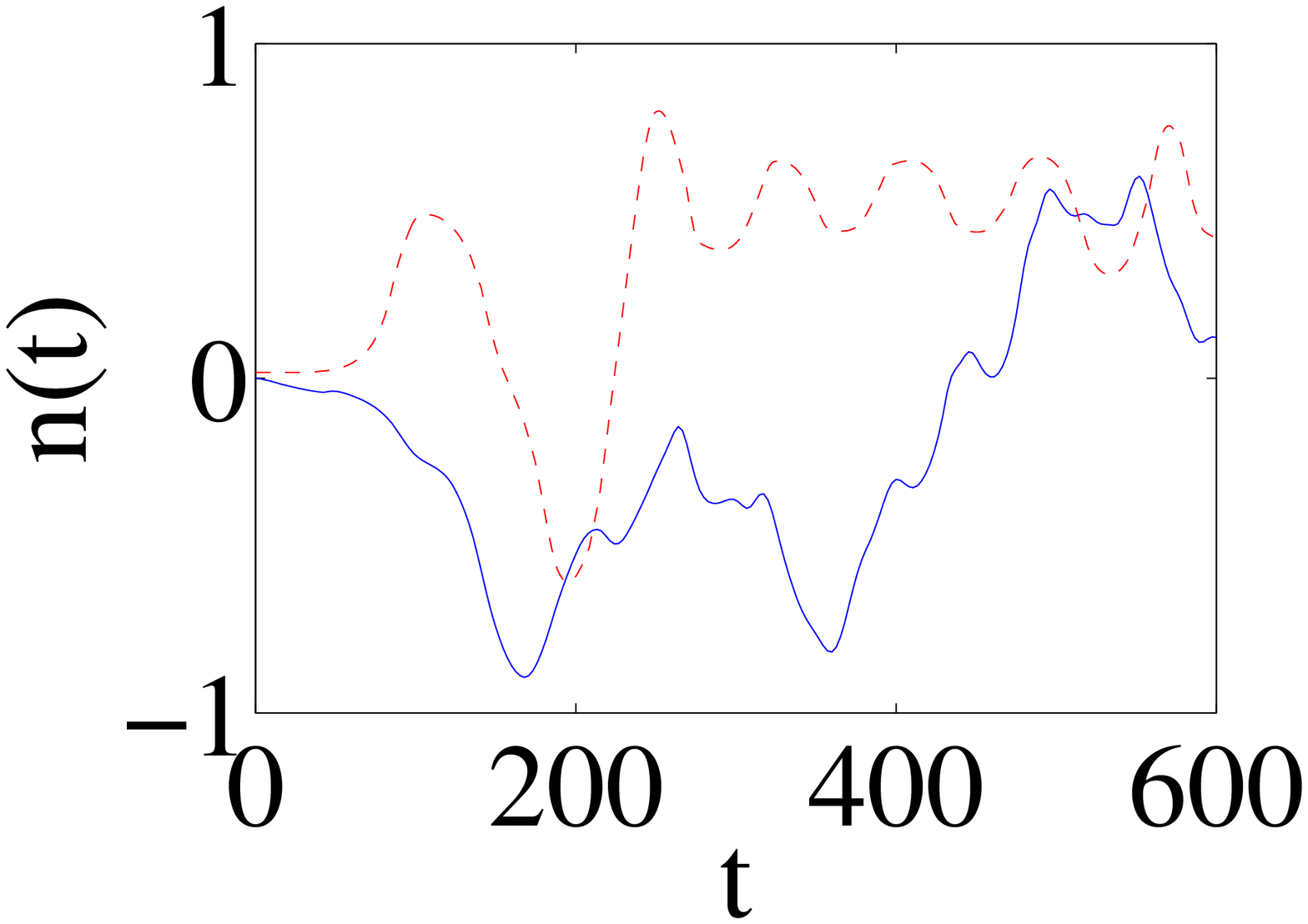}
\caption{(Color online) The top panels show the density contour plots and
the bottom panels the oscillations of the population imbalance between the
two potential wells, $n(t)$, in the case of the repulsive nonlinearity with $%
g_{0}=1$, $g_{1}=1.2$, and initial value of $\left( |\protect\psi %
|^{2}\right) _{\max }=0.1$. Left panel: $\protect\omega =0.005$; middle
panel: $\protect\omega =0.025$; right panel: $\protect\omega =0.15$. The
parameters of the potential are the same as in the previous figures. 
In the bottom panels,
the solid and dashed lines represent the results obtained from the
integration of the GPE and ODEs, respectively.}
\label{mu022_g12}
\end{figure}

Increasing the value of $g_{1}$, we observed qualitatively the same results
up to $g_{1}=0.9$, i.e., as long as the FRM amplitude, $g_{1}$, remained
smaller than the average value of the repulsive-nonlinearity coefficient, $%
g_{0}=1$. For $g_{1}>g_{0}$, when the total nonlinearity coefficient \emph{%
periodically changes its sign}, the situation is qualitatively different. In
particular, for $g_{1}=1.2$, the SSB (trapping in one well) occurs at $%
0.002\leq \omega <0.011$ and $0.088\leq \omega <0.097$ [see an example in
the left panels of Fig. \ref{mu022_g12}, for $\omega =0.005$]. At $0.011\leq
\omega <0.088$, the condensate remains untrapped [see middle panels of Fig. %
\ref{mu022_g12}, for $\omega =0.025$], while at $\omega >0.097$ the
oscillations of $n(t)$ are irregular, featuring neither symmetric
oscillations nor the self-trapping [see right panels of Fig. \ref{mu022_g12}%
, for $\omega =0.15$]. The parametric-resonance relation, $\omega _{\mathrm{%
trap}}\approx \omega /2$, is valid only in the interval of $0.088\leq \omega
<0.097$. 

\begin{figure}[tbh]
\includegraphics[width=8cm]{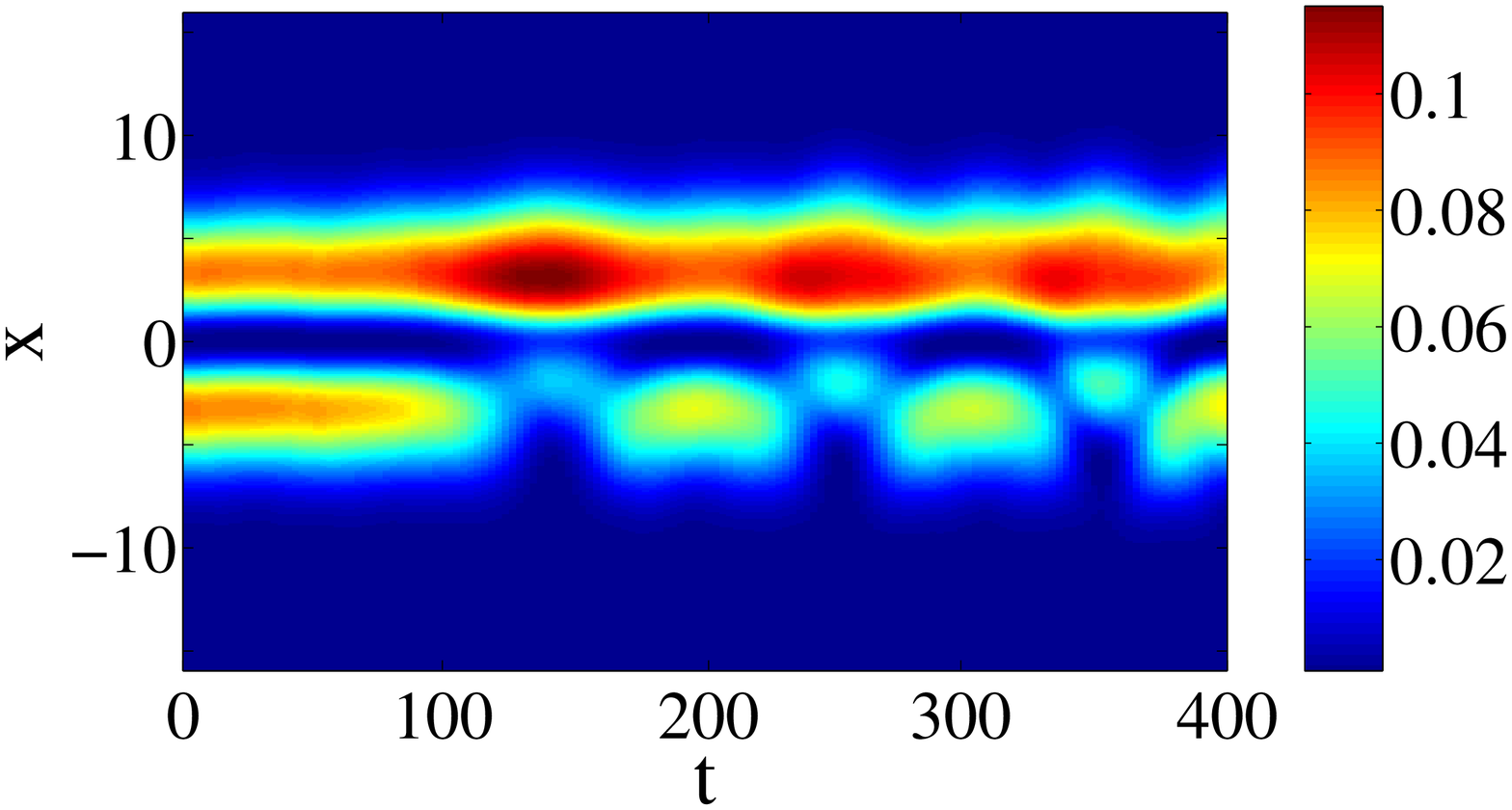} %
\includegraphics[width=8cm]{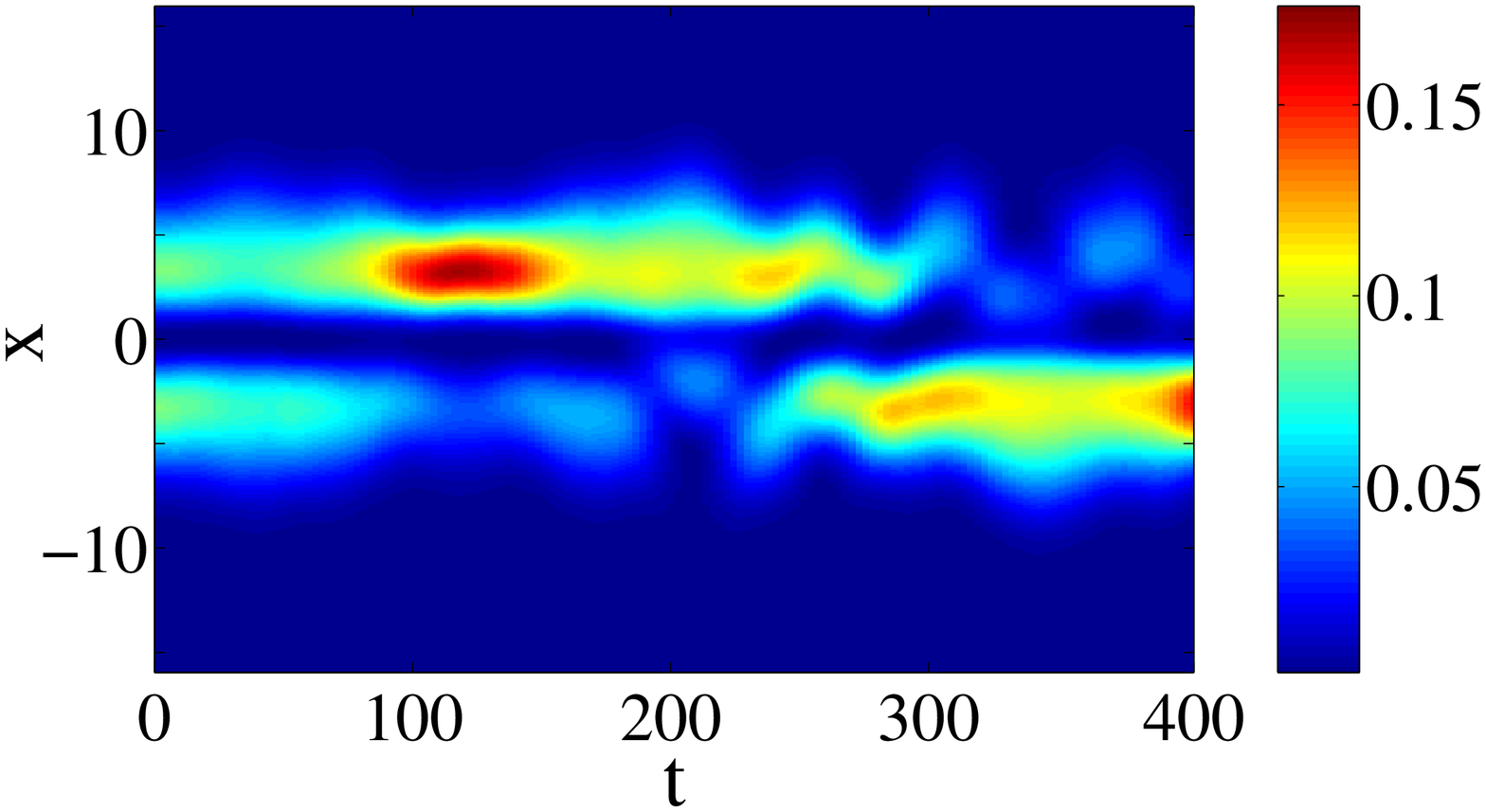} %
\includegraphics[width=8cm]{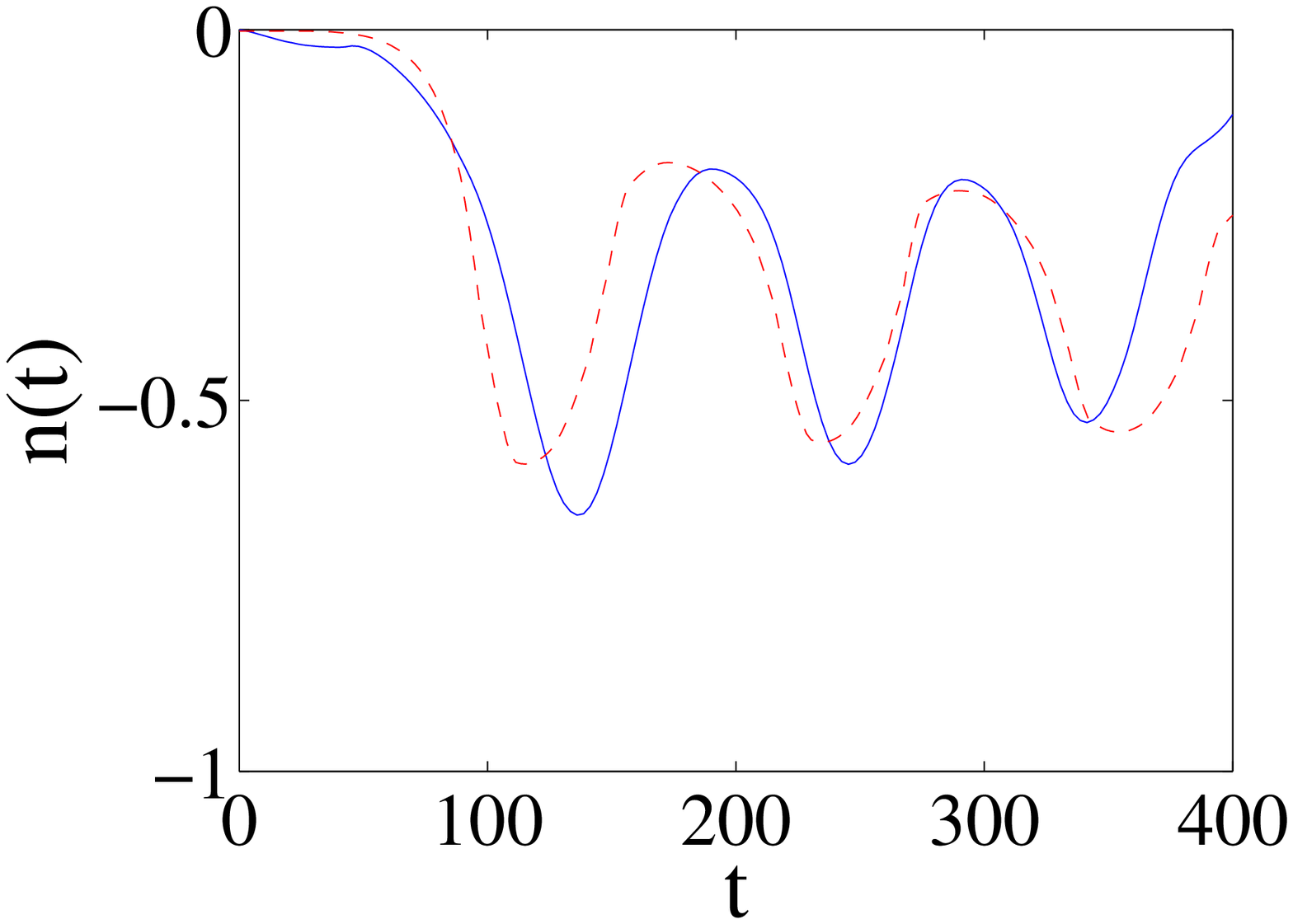} %
\includegraphics[width=8cm]{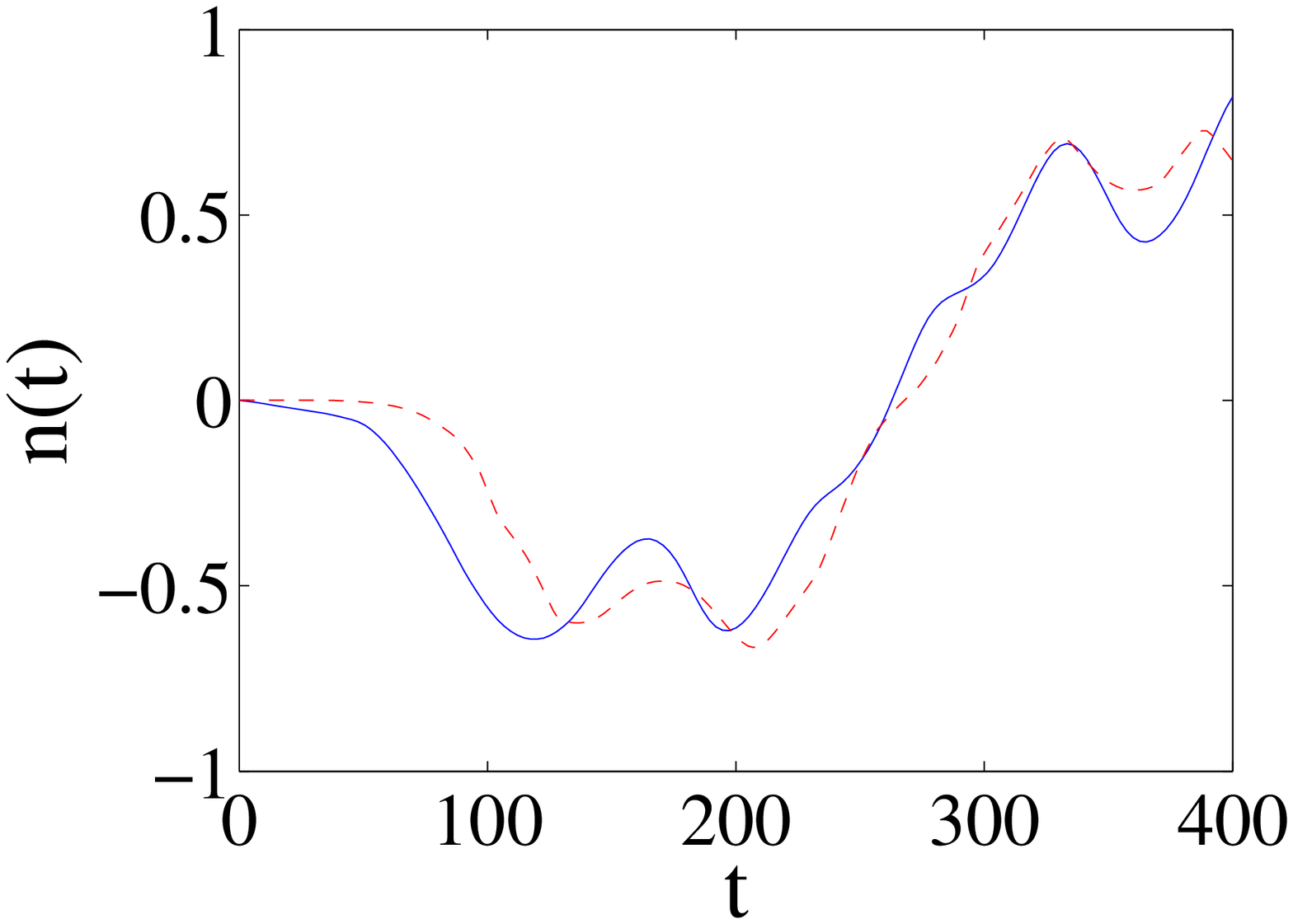}
\caption{(Color online) The evolution of the density contour plots (top) and
of $n(t)$ (bottom) in the case of the repulsive nonlinearity with $g_{0}=1$,
$g_{1}=1.5$, and initial value $\left( |\protect\psi |^{2}\right) _{\max
}=0.1$. Left and right panels pertain to $\protect\omega =0.005$ and $%
\protect\omega =0.055$, respectively. The potential is the same as in the
cases considered above. Additionally, In the bottom panels,
the solid and dashed lines represent the results obtained from the
integration of the GPE and ODEs, respectively.}
\label{mu022_g15}
\end{figure}

Similar results were obtained for $g_{1}=1.5$. In this case, the SSB
trapping occurs in the interval of $0.001\leq \omega <0.012$ [see left
panels of Fig. \ref{mu022_g15}, for $\omega =0.005$]. At $\omega >0.012$, $%
n(t)$ again features the irregular evolution, see an example in the right
panels of Fig. \ref{mu022_g15}, for $\omega =0.055$.
%In this case, relation $\omega_{{\rm osc}}\approx \omega/2$ does not hold at all.
Note that in these cases, corresponding to relatively large value of $g_1$, the 
Galerkin-type approximation for $n(t)$ fails to follow the result obtained in the framework of the GPE. This may be natural to expect as the large
strength of the nonlinearity renders relevant the inclusion
of additional modes in the description of the BEC dynamics.

%\begin{figure}[tbh]
%\includegraphics[width=5cm]{rep_m0022_w0005_c.eps} %
%\includegraphics[width=5cm]{rep_m0022_w0025_c.eps} %
%\includegraphics[width=5cm]{rep_m0022_w0150_c.eps}
%\caption{(Color online) Contour plots of the density in the case of the
%repulsive nonlinearity, with $g_{0}=1$, $g_{1}=1.2$ and initial values $|%
%\protect\psi |^{2}=0.1$. The panels pertain to the same values of $\protect%
%\omega $ as in Fig. \protect\ref{mu022_g12}.}
%\label{mu022_g12_c}
%\end{figure}

%\begin{figure}[tbh]
%\includegraphics[width=8cm]{rep_m0022_w0005_g115_c.eps} %
%\includegraphics[width=8cm]{rep_m0022_w0055_g115_c.eps}
%\caption{(Color online) Contour plots of the density for the same cases as
%in Fig. \protect\ref{mu022_g15}.}
%\label{mu022_g15_c}
%\end{figure}

\subsection{``Large'' initial values of the peak density}

\begin{figure}[tbh]
\includegraphics[width=8cm]
{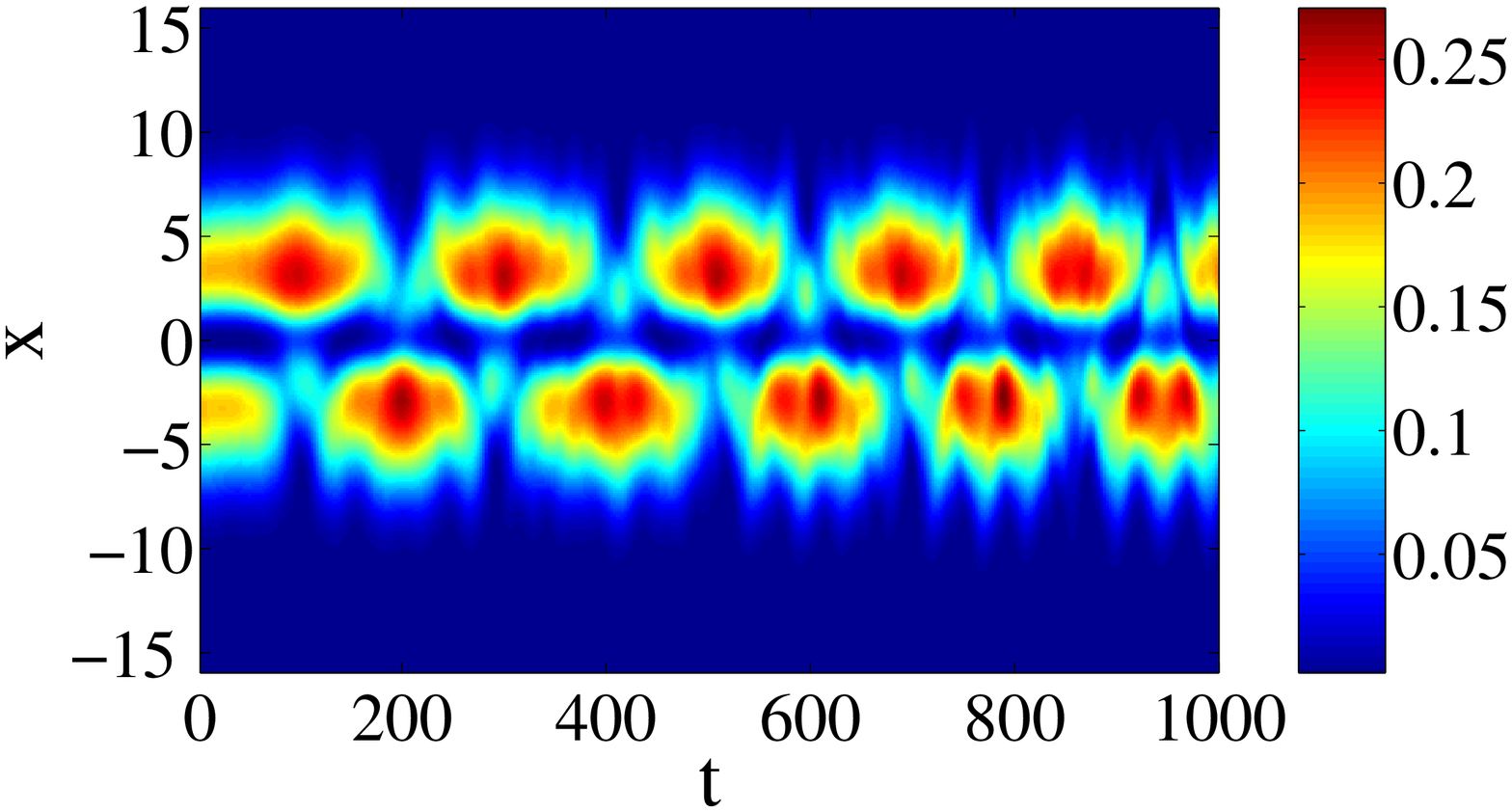}
\includegraphics[width=8cm]
{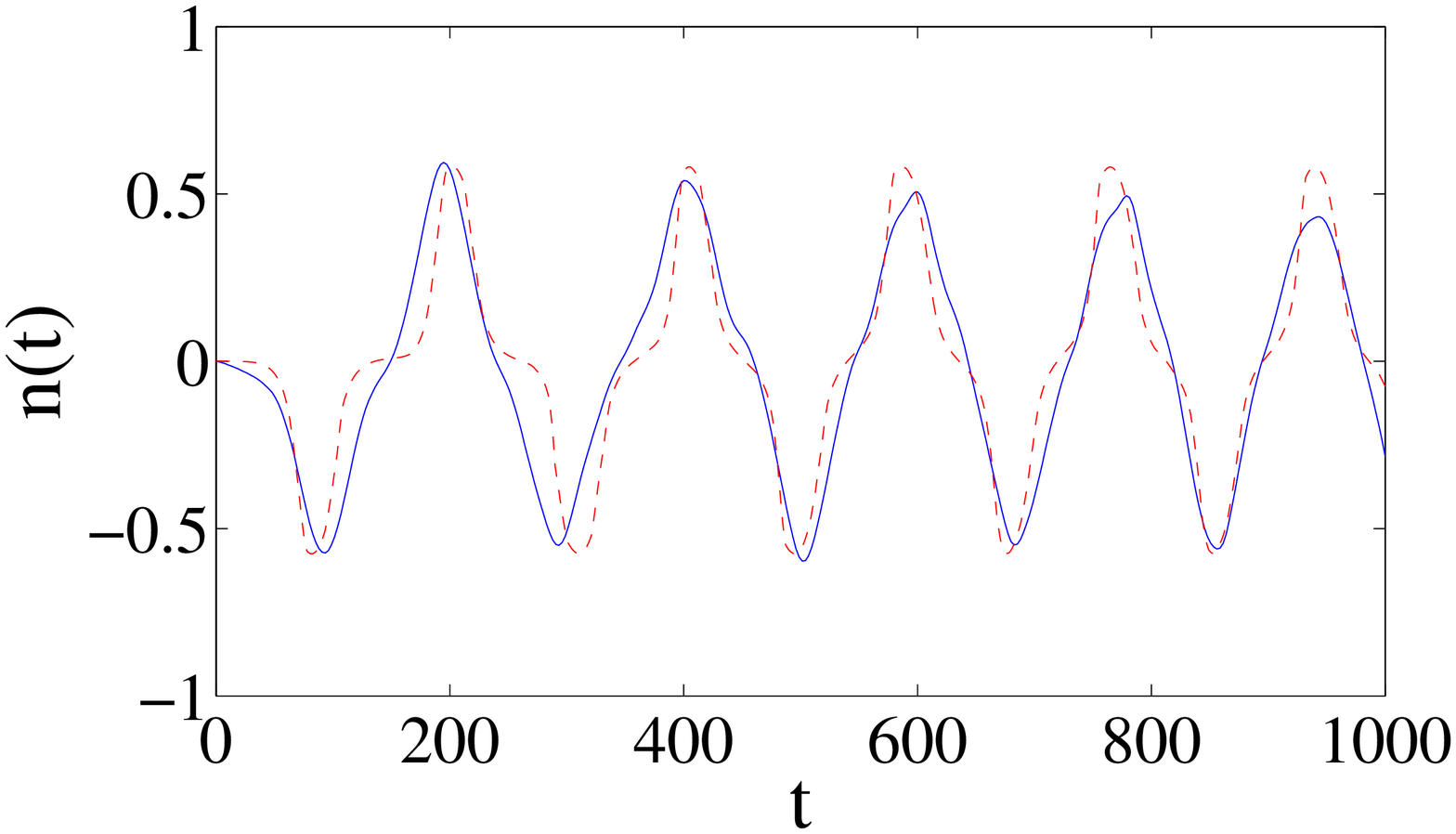}%{figure_017_g_0_traj.eps}
\caption{(Color online) The same as in Fig. \protect\ref{fig1}, but for the
initial value $\left( |\protect\psi |^{2}\right) _{\max }=0.17$. In the right panel,
the solid and dashed lines represent the results obtained from the
integration of the GPE and ODEs, respectively. The parameters
of the potential  are the same as in the previous case.}
\label{fig3}
\end{figure}

Here, we extend the analysis, increasing the initial values to $\left( |\psi
|^{2}\right) _{\max }=0.17$, while all the other DWP parameters are the same
as above (i.e. $\Omega =0.1$ and $V_{0}=1$). For $g_{1}=0$ (no management),
the results are shown in Fig. {\ref{fig3}}. In this case, the frequency of
the free oscillations of the condensate between the two wells is $\omega _{%
\mathrm{osc}}\approx 0.031$. The maximum value of the relative population
imbalance, $n(t)$, is $0.602$, and for larger values of the initial peak
density this value decreases.
% Bearing this criterion in mind [i.e. that the
%value of $n(t)$ be larger than $0.6$], we choose to investigate the system's
%behavior with this initial peak density.

\begin{figure}[tbh]
\includegraphics[width=8.4cm]
{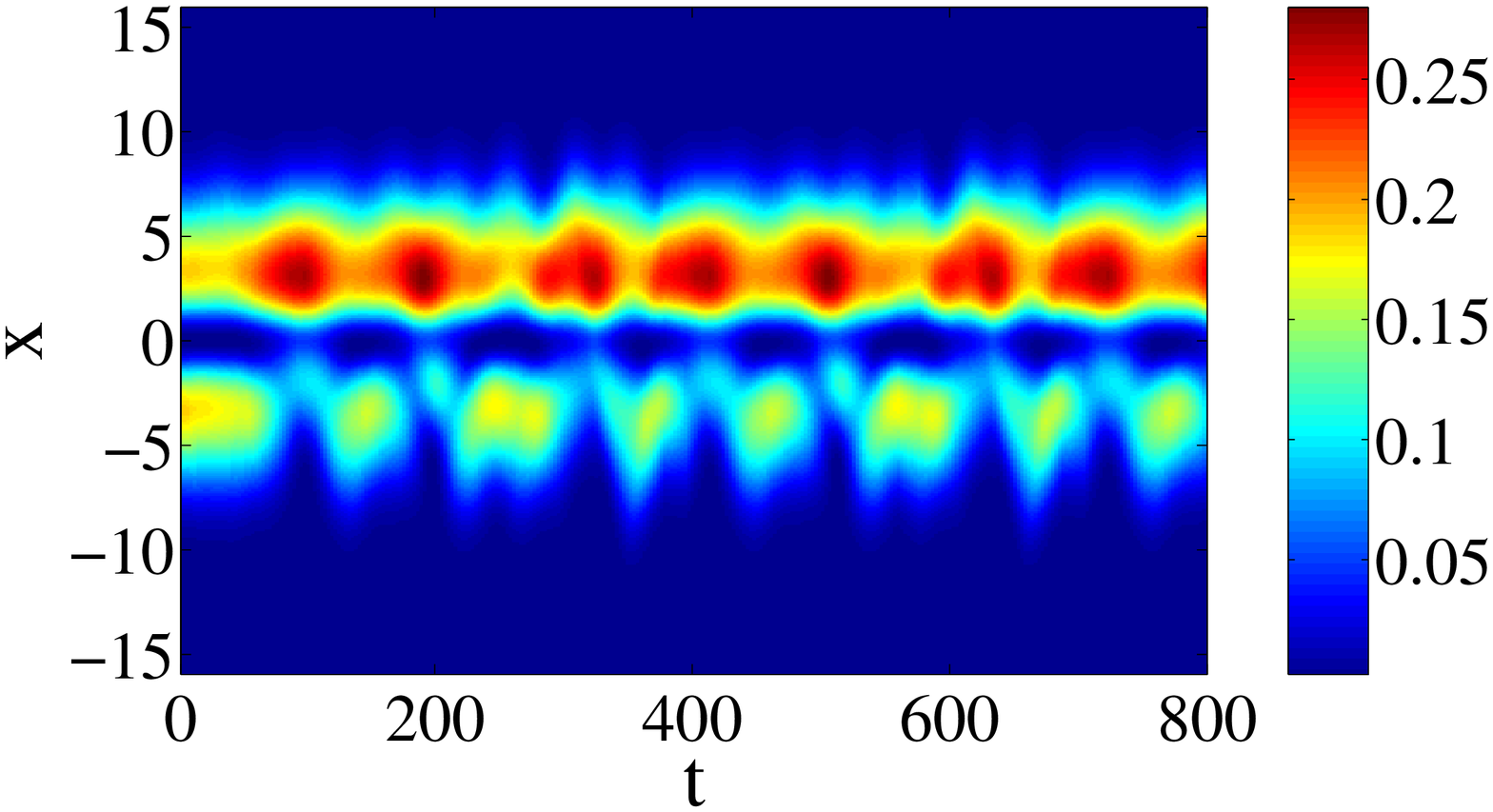}%{figure_017_g02_cont.eps}
\includegraphics[width=7.0cm]
{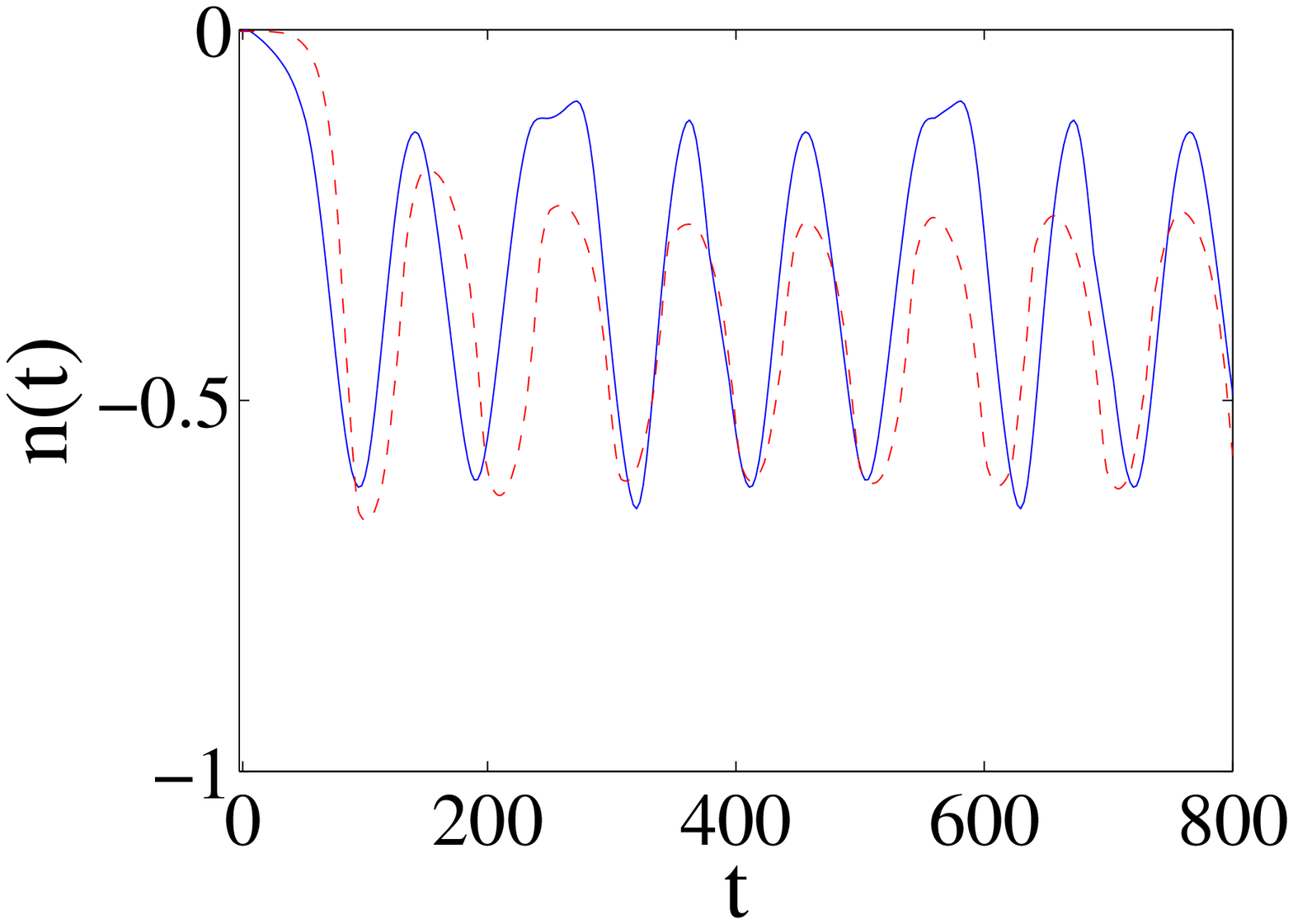}%{tr_rep_g1_02_om_008_large.eps}%{figure_017_g02_traj.eps}
\caption{(Color online) The same as in Fig. \protect\ref{fig2}, but for $%
g_{1}=0.2$, $\protect\omega =0.08$, and initial value $\left( |\protect\psi %
|^{2}\right) _{\max }=0.17$. In the right panel,
the solid and dashed lines represent the results obtained from the
integration of the GPE and ODEs, respectively. The other parameters are as in Fig. \protect\ref%
{fig2}.}
\label{fig4}
\end{figure}

Next, we study the behavior of $n(t)$ for different values of $g_{1}$ and $%
\omega $. For $g_{1}=0.2$, the result is that the condensate remains
untrapped at $0\leq \omega <0.06341$ and $\omega >0.1009$ [similar to the
situation displayed in Fig. {\ref{fig3}}], featuring the same oscillation
frequency as in the absence of the management, $\omega _{\mathrm{osc}%
}\approx 0.031$. For $0.06341\leq \omega <0.1009$ (i.e., $2\omega _{\mathrm{%
osc}}\lesssim \omega \lesssim 3\omega _{\mathrm{osc}}$), the evolution of $%
n(t)$ suggests trapping in one well, demonstrating the SSB effect. This
result is shown in Fig. \ref{fig4} for $\omega =0.08$. Notice that, 
as observed in the right panels of Figs.~\ref{fig3} and \ref{fig4}, 
the agreement between the ODEs and the GPE results is fairly good. 

Here, we should recall that, in the
previous case, with the smaller density peaks, the observed frequency of the
oscillations of the trapped BEC was following the parametric-resonance
relation, $\omega _{\mathrm{trap}}=\omega /2$. As the initial density peak
is getting larger, and, in particular, in the present case it corresponds to
$\left( |\psi |^{2}\right) _{\max }=0.17$, this rule is no longer valid. In
particular, the frequency of the trapped oscillations in the case of $\omega
=0.08$ is $0.0604$.

Next, we increase the FRM amplitude, $g_{1}$. For $g_{1}=0.4$, we have found
that, at $0\leq \omega <0.0598$ and $\omega >0.0904$, the trajectory of $%
n(t) $ remains untrapped, with the same oscillation frequency as in the
absence of the management. At $0.0598\leq \omega <0.0904$ (i.e., $2\omega _{%
\mathrm{osc}}\lesssim \omega \lesssim 3\omega _{\mathrm{osc}}$), the
oscillations of $n(t)$ feature the self-trapping in one well.

%\begin{figure}[tbh]
%\includegraphics[width=8cm]{traj02_m05_new.eps} %
%\includegraphics[width=8cm]{traj04_m05_new.eps}
%\caption{(Color online) The same as in Fig. \protect\ref{traj2}, but in the
%case when the initial values are $\left( \left\vert \protect\psi \right\vert
%^{2}\right) _{\max }=0.5$. Left panel: $g_{1}=0.2$, $0.07\leq \protect\omega %
%<0.11$. Right panel: $g_{1}=0.4$, $0.08\leq \protect\omega <0.13$.
%Parameters of the potential are $\Omega =0.1$ and $V_{0}=2$, other
%parameters being as in Fig. \protect\ref{traj2}.}
%\label{traj3}
%\end{figure}

\begin{figure}[tbh]
\includegraphics[width=8cm]
{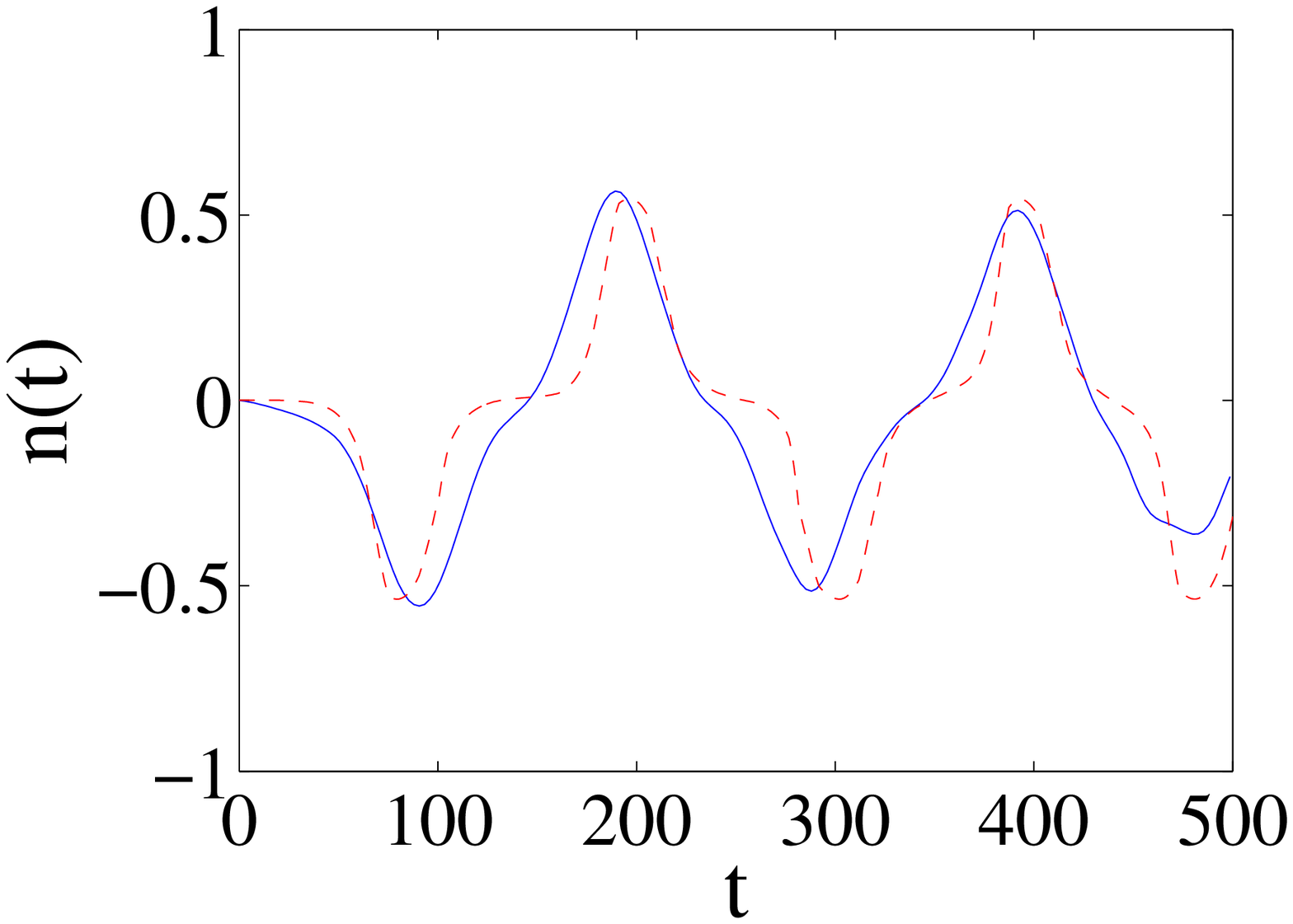}
\includegraphics[width=8cm]
{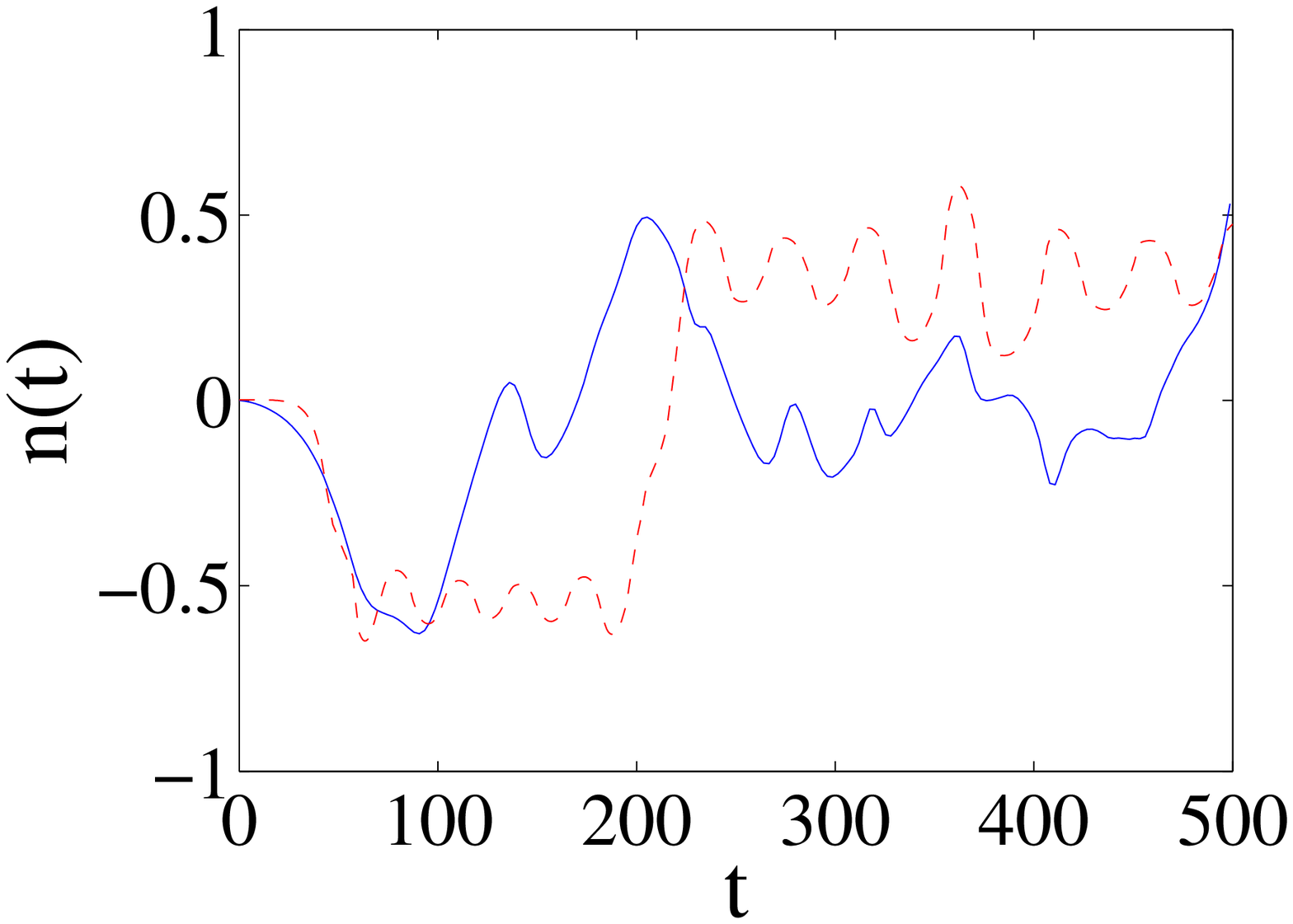}
\caption{(Color online) Trajectories of $n(t)$ in the case of the repulsive
nonlinearity with $g_{0}=1$, $g_{1}=1.2$, and initial value $\left( |\protect%
\psi |^{2}\right) _{\max }=0.17$. Left and right panels pertain,
respectively, to $\protect\omega =0.01$ and $\protect\omega =0.1$ with the
same potential parameters as above. Moreover, the solid and dashed lines 
represent the results obtained from the
integration of the GPE and ODEs, respectively.}
\label{mu05_g12}
\end{figure}

Qualitatively the same behavior is observed for $g_{1}<0.95$. At greater
values of $g_{1}$, we did not observe any self-trapping. More specifically,
we examined the same two values of the FRM amplitude as above, i.e., $%
g_{1}=1.2$ and $g_{1}=1.5$. For $g_{1}=1.2$, we observed untrapped periodic
oscillations for $0\leq \omega <0.0561$ [see the left panel of Fig. \ref{mu05_g12} 
%and the
%left panel in Fig. (\ref{mu05_g12_c}),
for $\omega =0.01$], while for $\omega >0.02$ the population imbalance, 
$n(t)$, features an irregular evolution, 
see the right panel of Fig. \ref{mu05_g12} 
%and the right panel in
%Fig. (\ref{mu05_g12_c}),
for $\omega =0.1$. Notice that, again, the agreement between the ODEs 
and GPE result is less good for this case of large nonlinearity due
to the apparent involvement of higher modes in the dynamics.

Similar results were obtained for $g_{1}=1.5$. In this case, untrapped
periodic oscillations occur at $0\leq \omega <0.001$, resembling the
situation displayed in the left panel of Fig. \ref{mu05_g12}, while, at $%
\omega >$ $0.001$, irregular evolution of $n(t)$ is again observed, similar
to that in the right panel of Fig. \ref{mu05_g12}.

%\begin{figure}[htb]
%\includegraphics[width=8cm]{rep_m05_w001_g112_c.eps} %
%\includegraphics[width=8cm]{rep_m05_w003_g112_c.eps}
%\caption{Contour plots for the case of repulsive nonlinearity with $g_0=1$, $%
%g_1=1.2$ and initial $|\protect\psi|^2=0.5$. Left panel: $\protect\omega%
%=0.01 $, Right panel: $\protect\omega=0.03$. The potential parameters are $%
%\Omega=0.1$, $V_0=2$, $x_0=0$ and $w=0.5$.}
%\label{mu05_g12_c}
%\end{figure}
%\textbf{[I HAVE COMMENTED OUT FORMER FIG. 12 AND REFERENCES TO IT IN THE
%TEXT, AS THIS FIGURE SEEMS REDUNDANT.]}

In Table I, we summarize intervals of $g_1/\omega$ in which the
trapped oscillations take place, in the cases of smaller and larger
initial values of the peak density $(|\psi|^2)_{max}$ [i.e., $0.1$
and $0.17$], for $g_1=0.2$ and $0.4$. In these cases, we observe
that the trapped oscillations appear at values of $g_1/\omega$
similar to those presented in Ref. \cite{xie}, in the framework of
the two-mode approximation. However, the similarity to the two-mode
approximation is lost at larger values of $g_1$.

\begin{table}[ht]
\label{tabl1} \caption{Intervals of $g_1/\omega$ for trapped
oscillations}
\begin{tabular} {|c|c|c|}
\hline
 Initial $(|\psi|^2)_{max}$ & $g_1$ & $g_1/\omega$ (interval of trapped oscillations)  \cr
\hline
 $0.1$& $0.2$ &$1.193 \leq g_1/\omega \leq 2.522$ \cr
 & $0.4$ &$2.530 \leq g_1/\omega \leq 4.635$ \cr
\hline
 $0.17$& $0.2$ &$1.982 \leq g_1/\omega \leq 3.154$ \cr
 & $0.4$ &$4.424 \leq g_1/\omega \leq 6.689$ \cr
 \hline
\end{tabular}
\end{table}

\section{The attractive nonlinearity}

\begin{figure}[tbh]
\includegraphics[width=8cm]{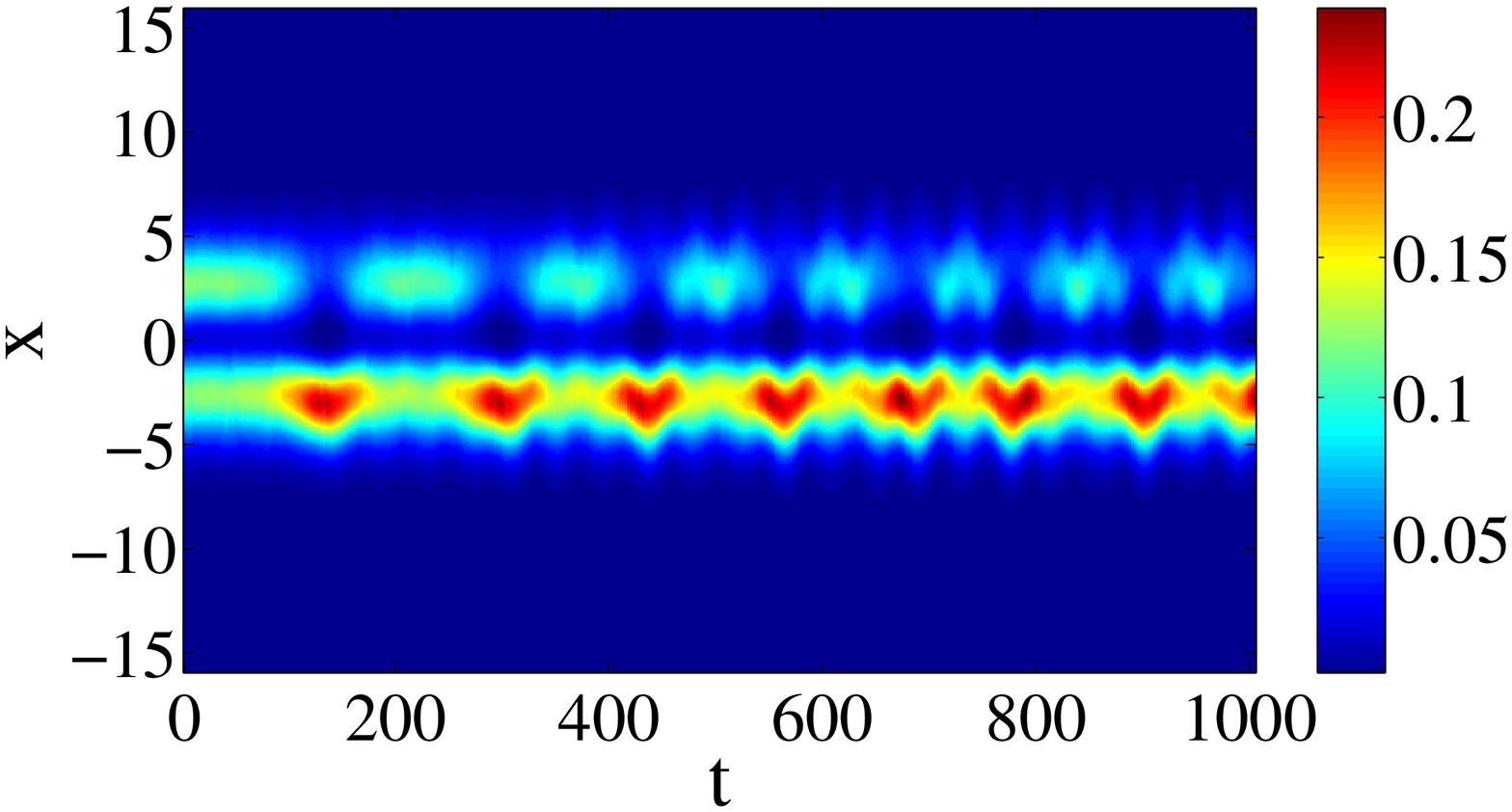} %
\includegraphics[width=8cm]{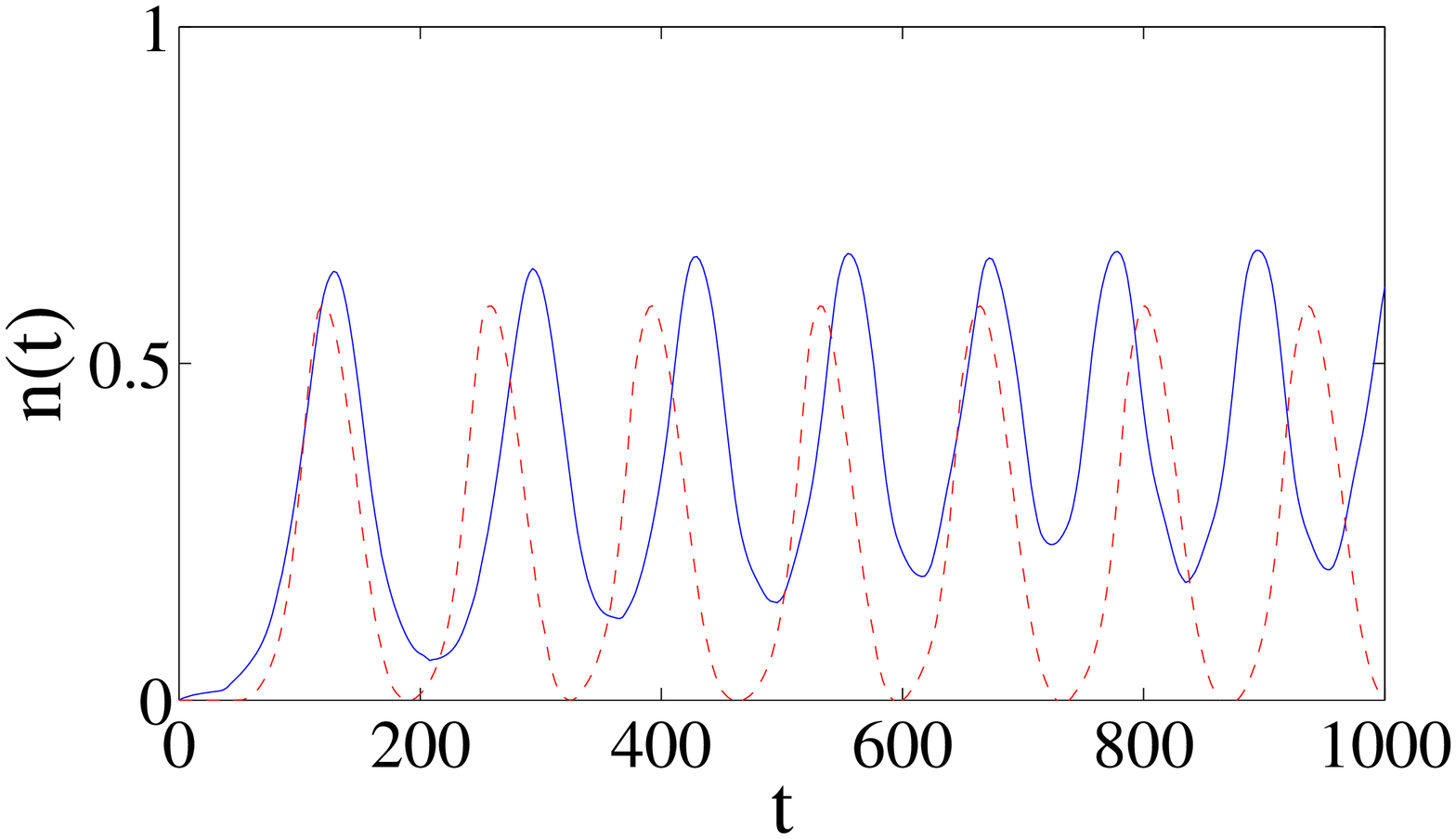}%{t1000_traj.eps}
\caption{(Color online) The same as in Fig. \protect\ref{fig1}, but in the
case of the attractive nonlinearity, with $g_{0}=-1$, $g_{1}=0$, and initial
value $\left( |\protect\psi |^{2}\right) _{\max }=0.13$. In the right panel,
the solid and dashed lines represent the results obtained from the
integration of the GPE and ODEs, respectively. This figure and
others in this section correspond to the same parameters of the potential as
above, i.e., $\Omega =0.1$, $V_{0}=1$, and $w=0.5$.}
\label{fig5}
\end{figure}

In the version of the DWP model with the self-attraction, we start the
simulations, without the application of the FRM, for values of the
parameters similar to those of Ref. {\cite{theo}}: $g_{0}=-1$, $g_{1}=0$, $%
\Omega =0.1$, $V_{0}=1$, $w=0.5$, and the initial value of $\left( |\psi
|^{2}\right) _{\max }=0.13$. Results of the simulations are shown in Fig. {%
\ref{fig5}}. In this case, the unmanaged condensate is self-trapped in one
well, oscillating in it at a frequency $\omega _{\mathrm{trap,osc}}=0.046$.

\begin{figure}[tbh]
\includegraphics[width=8cm]{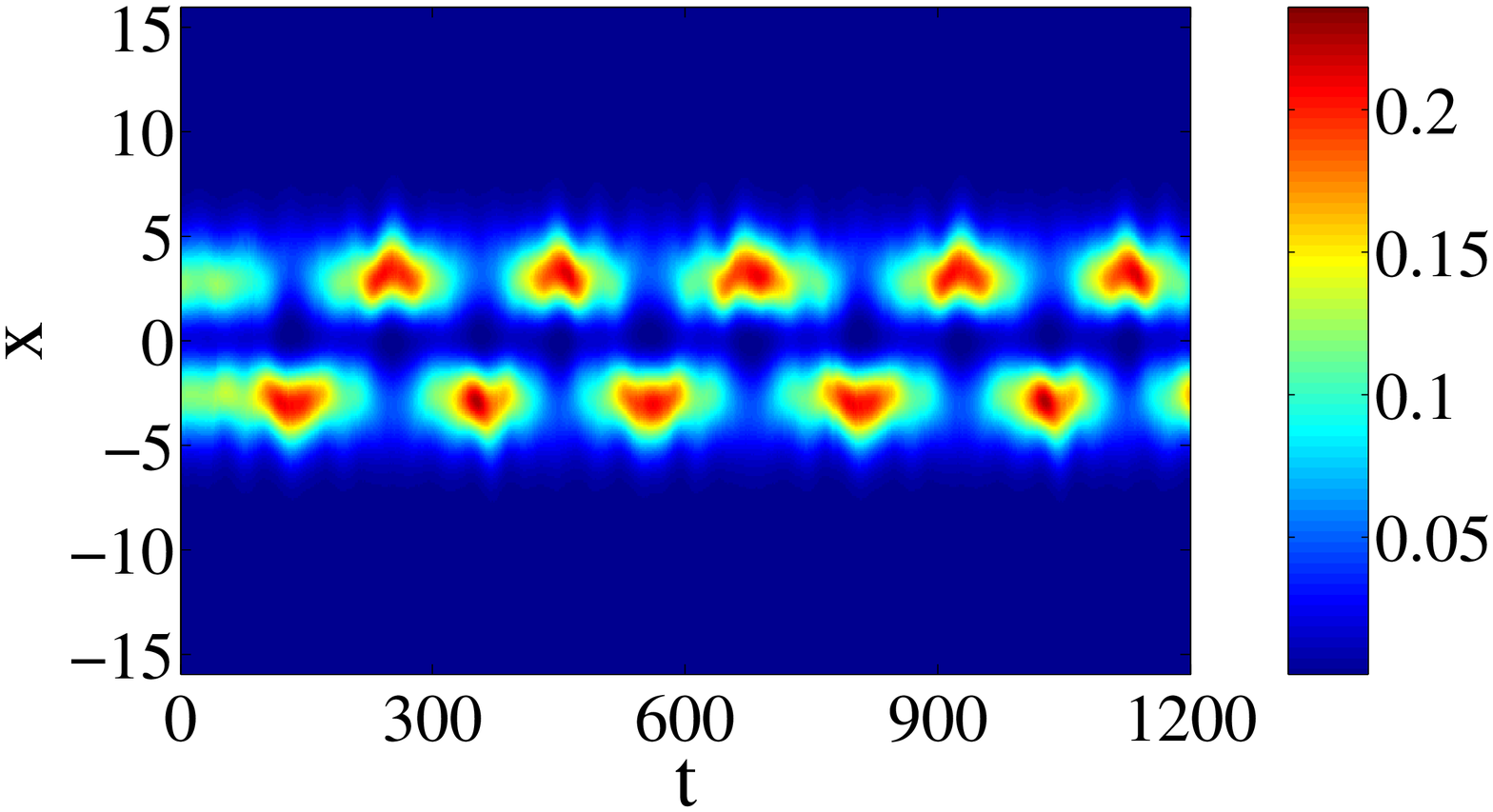}%{fig6a_new.eps} %
\includegraphics[width=8cm]{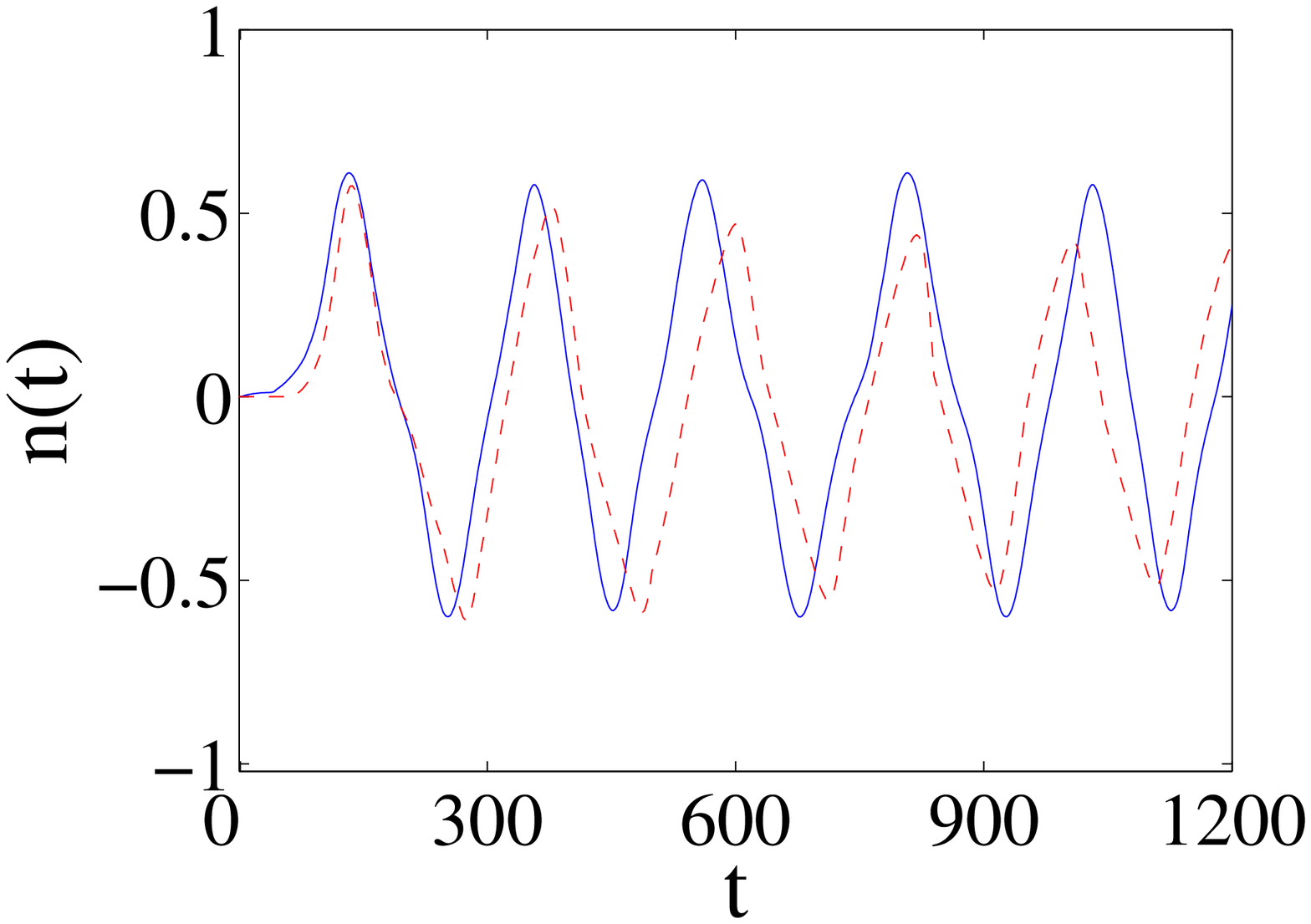}%{tr_attr_g1_02_om_013.eps}%{fig6b_new.eps}
\caption{(Color online) The same as in Fig. \protect\ref{fig2}, but in the
model with the attractive nonlinearity, $g_{0}=-1$, with $g_{1}=0.2$, $%
\protect\omega =0.13$, and initial value $\left( |\protect\psi |^{2}\right)
_{\max }=0.13$. In the right panel,
the solid and dashed lines represent the results obtained from the
integration of the GPE and ODEs, respectively.}
\label{fig6}
\end{figure}

In the case of the dominating attractive nonlinearity, the main effect of
the FRM is opposite to that in the model with the repulsion: \emph{untrapping%
} of the condensate which, in the unmanaged case, was trapped in the SSB
state. For $g_{1}=0.2$, we have found that the untrapping occurs at%
\begin{equation}
0.1241\leq \omega \leq 0.1594  \label{attract}
\end{equation}%
(or, almost exactly, $3\omega _{\mathrm{trap,osc}}\leq \omega <4\omega _{%
\mathrm{trap,osc}}$). This result is shown in Fig. \ref{fig6} for $\omega
=0.13$. In this case, the observed frequency of the oscillations of $n(t)$
in the FRM-induced untrapped state is $\omega _{\mathrm{untrap}%
}=0.0318\approx \omega /4$, suggesting that this may be a manifestation of a
higher-order parametric resonance \cite{resonance}. Notice that, 
as observed in the right panels of Figs.~\ref{fig5} and \ref{fig6}, 
the agreement between the ODEs and the GPE results is good.

\begin{figure}[tbh]
\includegraphics[width=8cm]{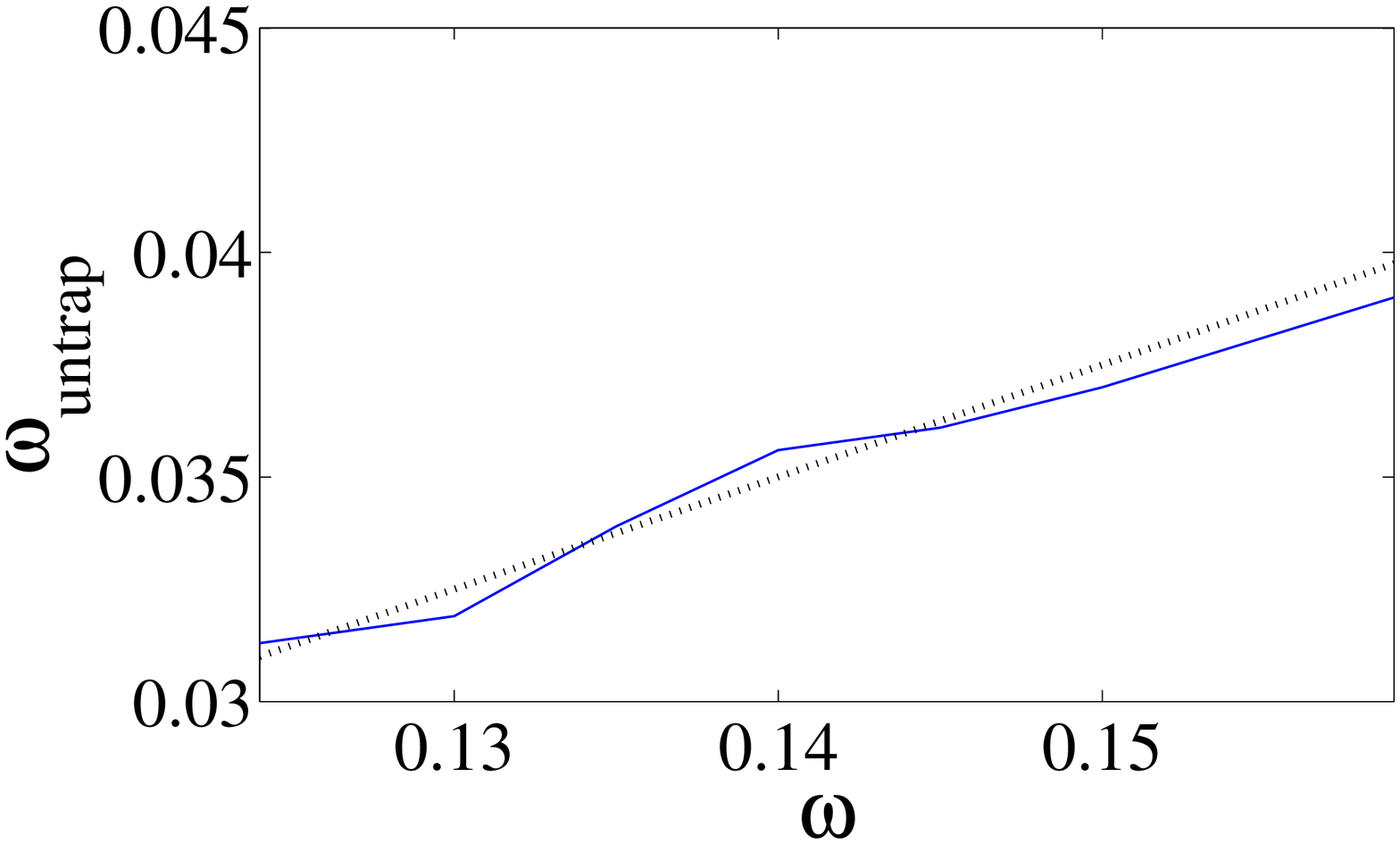}%{traj02.eps} %
\includegraphics[width=8cm]{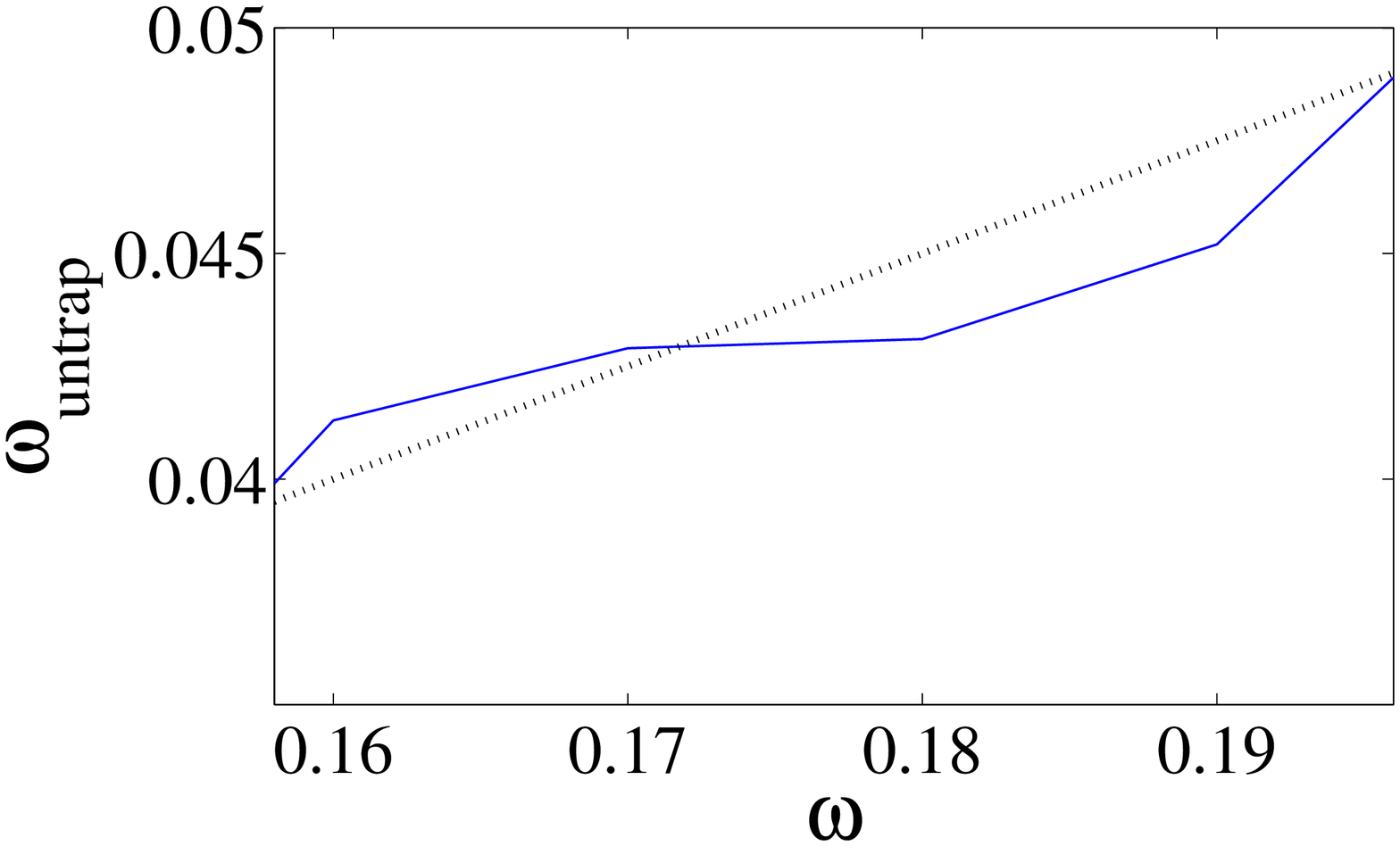}%{traj04.eps}
\caption{(Color online) The relation between $\protect\omega $ and $\protect%
\omega _{\mathrm{untrap}}$ in the model with the attractive nonlinearity.
Solid, and dotted lines correspond, respectively, to the result obtained
from the GPE, and the estimation of $\omega _{\mathrm{untrap}}= \omega /4$.
Left panel: $g_{0}=-1$, $g_{1}=0.2$, $0.124\leq \protect\omega <0.159$, and
initial values $|\protect\psi |^{2}=0.13$. Right panel: $g_{0}=-1$, $%
g_{1}=0.4$, $0.158\leq \protect\omega <0.196$, and initial values $|\protect%
\psi |^{2}=0.35$.}
\label{traj4}
\end{figure}

{We studied the oscillations of the untrapped state for other values of $%
\omega $ from interval (\ref{attract}). It has been concluded that, at $%
\omega =0.12$ and }$0.14${, the observed frequency of the untrapped
oscillations again obeys the above-mentioned relation suggesting a
higher-order parametric resonance: $\omega _{\mathrm{untrap}}=0.0298\approx
\omega /4$, and $\omega _{\mathrm{untrap}}=0.0346\approx \omega /4$,
respectively. In the left panel of Fig. (\ref{traj4}), we present the
relation between $\omega $ and $\omega _{\mathrm{untrap}}$ in the region of $%
0.124\leq \omega <0.159$. For $\omega $}${<}${\ $0.124$ and }${\omega >}${$%
0.159$, the oscillations of $n(t)$ remain trapped in a single well, with the
corresponding frequency almost the same as in the absence of the management,
i.e., $\omega _{\mathrm{trap,osc}}=0.046$. }

Next, we increased the value of the FRM amplitude, $g_{1}$. For $g_{1}=0.4$,
the untrapping occurred at $0.158\leq \omega <0.196$ (which almost exactly
means $4\omega _{\mathrm{trap,osc}}\leq \omega <5\omega _{\mathrm{trap,osc}}$%
). As above, we examined the relation between $\omega $ and the respective
frequency of the untrapped oscillations, $\omega _{\mathrm{untrap}}$. The
results verify the same conclusion as reached above, i.e., {$\omega _{%
\mathrm{untrap}}\approx \omega /4$}. In the right panel of Fig. \ref{traj4},
we present the values of $\omega $ vs. $\omega _{\mathrm{untrap}}$ in the
region of $0.158\leq \omega <0.196$.

\begin{figure}[tbh]
\includegraphics[width=5cm]{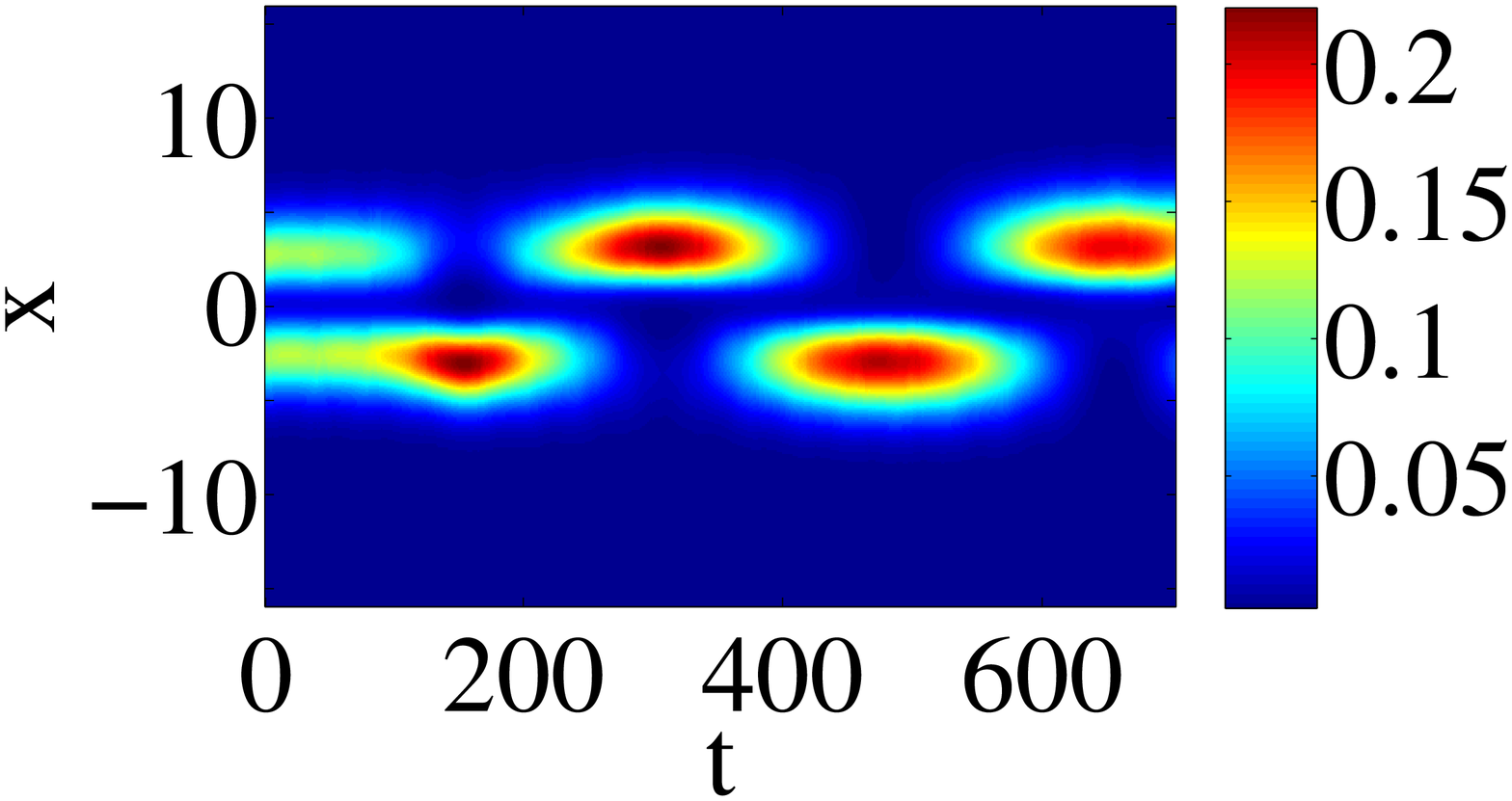} %
\includegraphics[width=5cm]{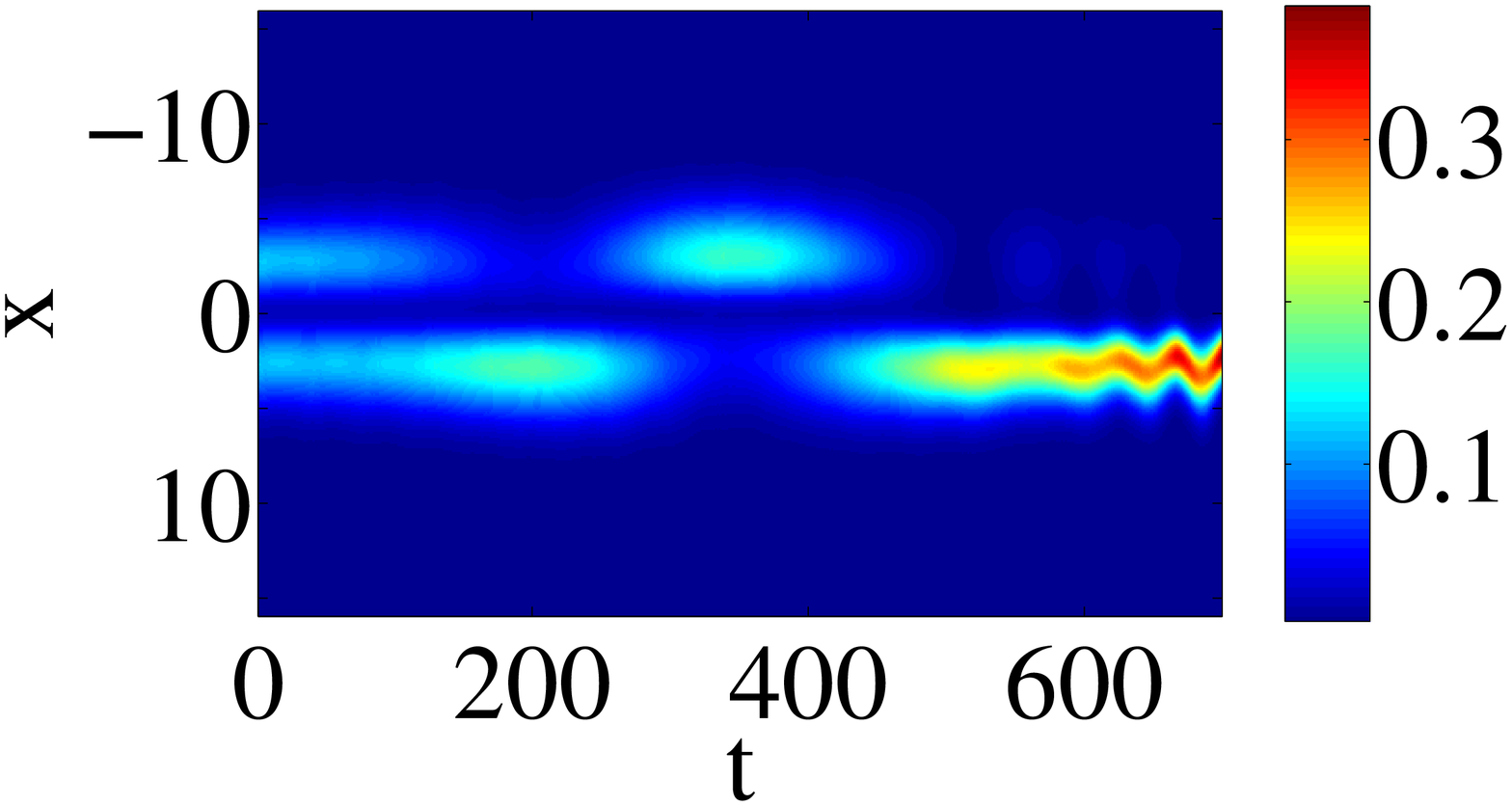} %
\includegraphics[width=5cm]{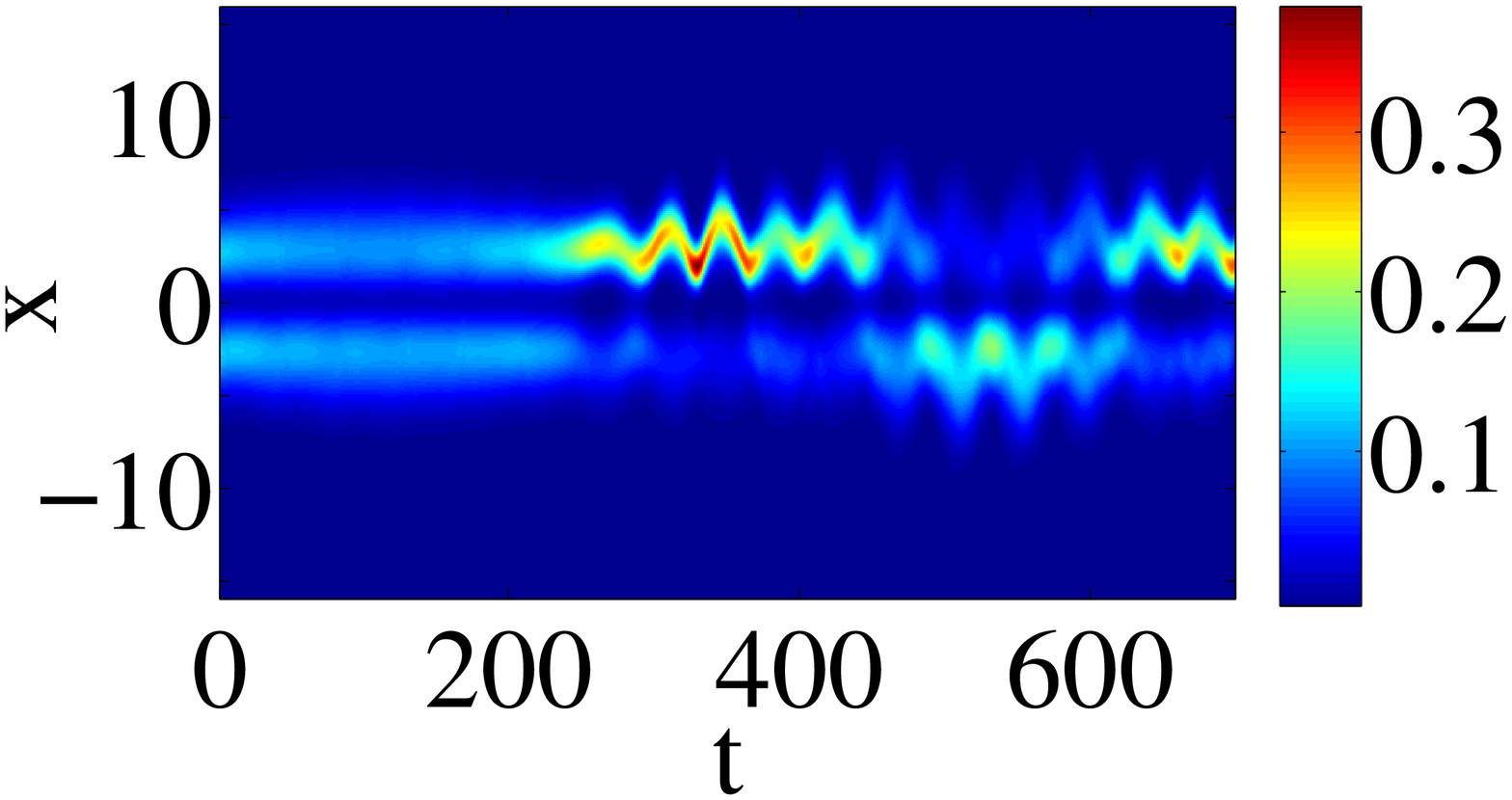} %
\includegraphics[width=5cm]{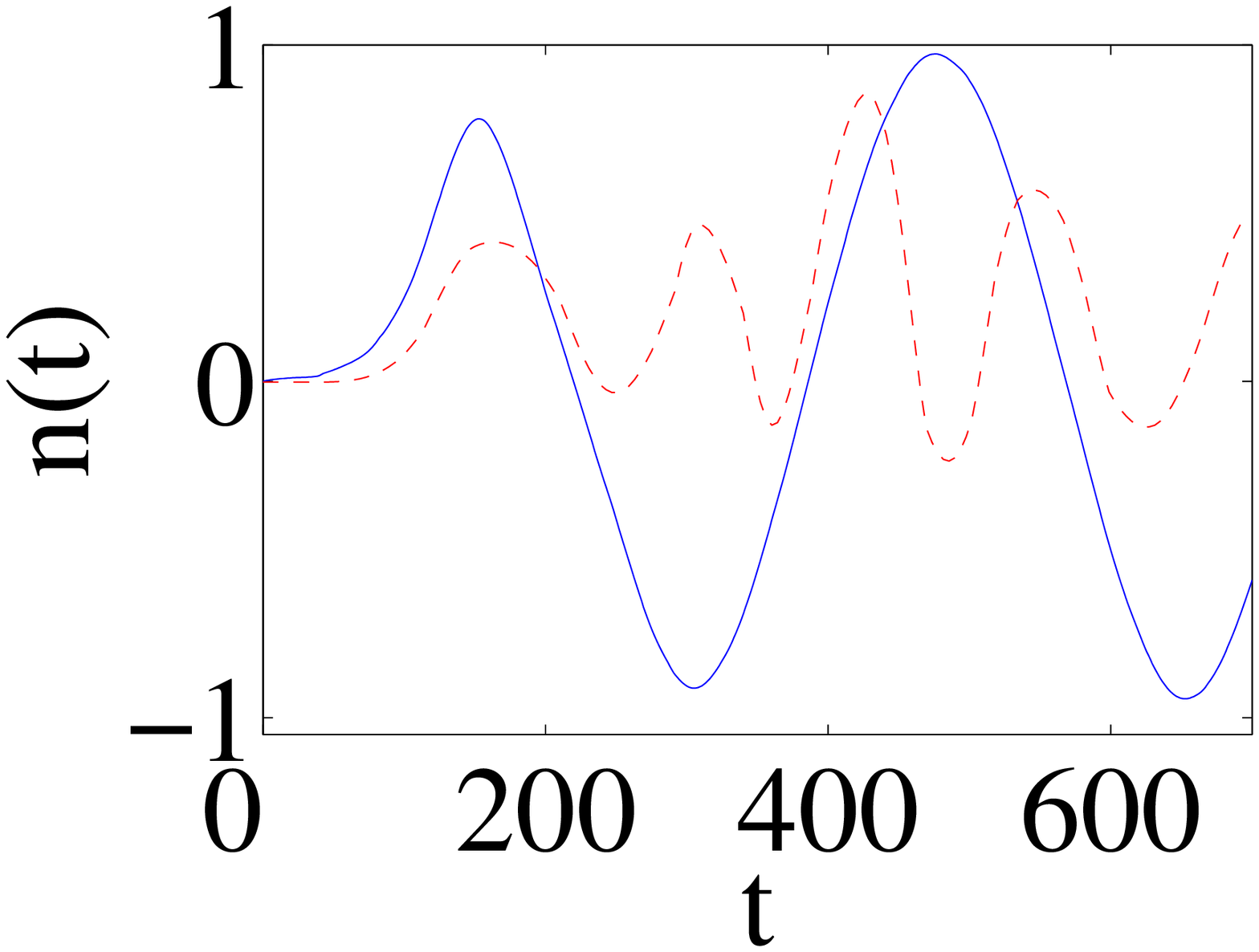} %
\includegraphics[width=5cm]{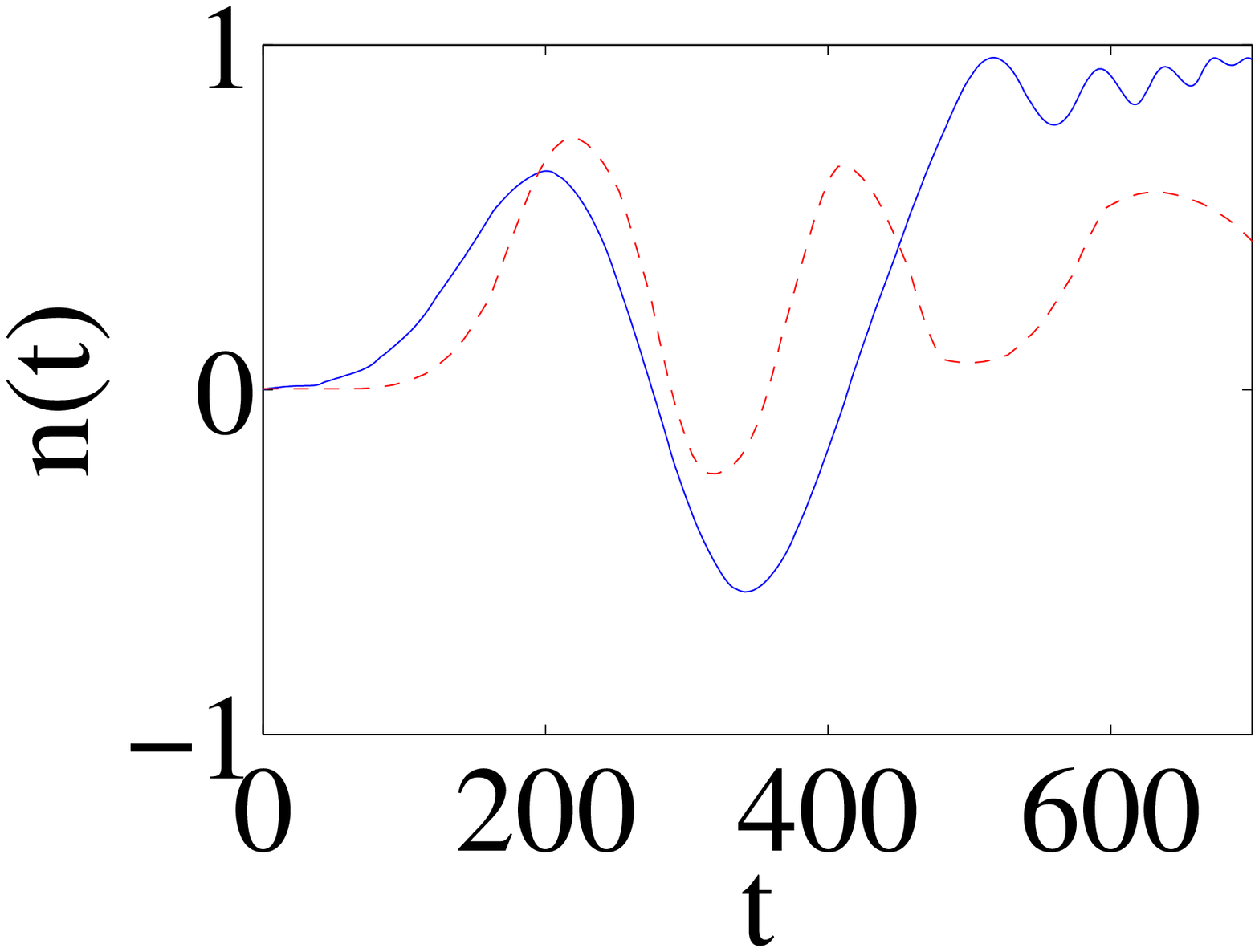} %
\includegraphics[width=5cm]{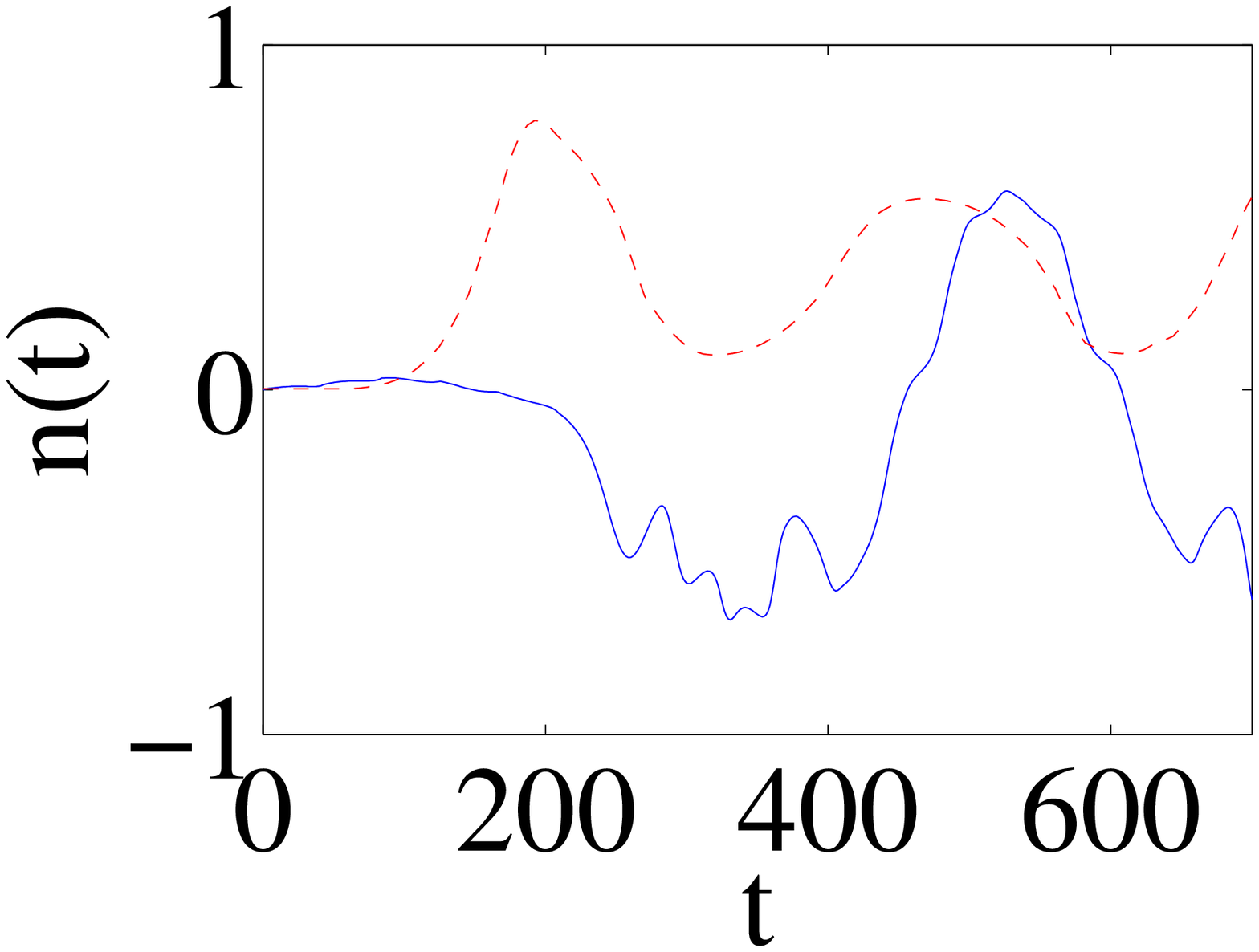}
\caption{(Color online) Spatiotemporal contour plots (top panels) and
trajectories of $n(t)$ (bottom panels) in the case of the attractive
nonlinearity with $g_{0}=-1$, $g_{1}=1.2$, and initial value $\left( |%
\protect\psi |^{2}\right) _{\max }=0.13$. Left panel: $\protect\omega =0.003$%
; middle panel: $\protect\omega =0.007$; right panel: $\protect\omega =0.02$. Additionally, In the bottom panels,
the solid and dashed lines represent the results obtained from the
integration of the GPE and ODEs, respectively.}
\label{mu005_g12}
\end{figure}

As in the case of the repulsive nonlinearity, we have also examined the
cases with $g_{1}>|g_{0}|$, when the total nonlinearity coefficient
periodically changes its sign. This was studied for $g_{1}=1.2$ and $%
g_{1}=1.5$. In the former case, at $0.001<\omega <0.006$ the population
imbalance, $n(t)$, oscillates in the untrapped state, see the left panels of
Fig. \ref{mu005_g12} %, and the left panel in
%Fig. (\ref{mu005_g12_c})
for $\omega =0.003$. At $0.007\leq \omega <0.014$, we observe re-trapping
after one or two untrapped oscillations, see the middle panels in Fig. \ref%
{mu005_g12} %and the middle panel in Fig. (\ref{mu005_g12_c}),
for $\omega =0.007$, while at $\omega >0.014$ the evolution of $n(t)$ is
irregular, as seen in the right panels of Fig. \ref{mu005_g12}
% and the right panel in Fig. (\ref{mu005_g12_c}),
for $\omega =0.02$. Similar results were obtained for $g_{1}=1.5$: in this
case, $n(t)$ oscillates without trapping at $0.001\leq \omega <0.004$.
%[similar to what is
%shown in Figs. (\ref{mu005_g12}) and the left panel of Fig. (\ref%
%{mu005_g12_c})].
For $0.004\leq \omega <0.012$, we observed re-trapping after one or two
oscillations, %[similar to Fig. (\ref{mu005_g12}) and the middle
%panel in Fig. (\ref{mu005_g12_c})],
and the evolution of $n(t)$ is irregular at $\omega >0.012$.
%, similar to the situation displayed in Fig. (\ref%
%{mu005_g12}) and the right panel of Fig. (\ref{mu005_g12_c}).
As observed in the bottom panels of Fig.~\ref{mu005_g12}, it is obvious that the 
results obtained from the ODEs and the GPE are not in agreement for values of $g_1>1$, for similar reasons as considered above.

%\begin{figure}[tbh]
%\includegraphics[width=5cm]{attr_m0005_w0003_c.eps} %
%\includegraphics[width=5cm]{attr_m0005_w0007_c.eps} %
%\includegraphics[width=5cm]{attr_m0005_w0020_c.eps}
%\caption{(Color online) Spatiotemporal contour plots of the density in the
%case of the attractive nonlinearity, with $g_{0}=-1$, $g_{1}=1.2$ and
%initial $\left( |\protect\psi |^{2}\right) _{\max }=0.13$. Left panel: $%
%\protect\omega =0.003$; middle panel: $\protect\omega =0.007$; right panel: $%
%\protect\omega =0.02$. Parameters of the potential are $\Omega =0.1$, $%
%V_{0}=1$, and $w=0.5$.}
%\label{mu005_g12_c}
%\end{figure}
%\textbf{[THE CAPTION TO THE LAST FIGURE IS OBVIOUSLY WRONG. PLEASE REPLACE
%IT BY AN APPROPRIATE ONE. THIS FIGURE SHOWS NOT TRAJECTORIES OF }$n(t)$%
%\textbf{\ BUT CONTOUR PLOTS OF THE DENSITY.]}

\section{Conclusions}

We have studied the trapping/detrapping of Bose-Einstein condensates
confined in a DWP (double-well potential), under the action of the FRM
(Feshbach-resonance-management) technique. Our model is based on the
Gross-Pitaevskii equation with the time-dependent nonlinear coefficient
whose average value corresponds to either repulsion or attraction. We have
found that the two signs give rise to different effects, regarding the
trapping/detrapping of the condensate in the DWP. In the case of the
self-repulsion, the BEC gets trapped in one of the two wells under the
action of the FRM, if it was untrapped in the unmanaged state. We have found
the range of values of the driving frequency leading to this effect. If the
sign of the nonlinearity does not change due to the time-periodic
modulation, 
and the densities are sufficiently low, as mentioned above, 
the 
frequency of the intrinsic oscillations of the FRM-induced trapped states 
is half the management frequency, which suggests a parametric resonance as 
an underlying mechanism.

On the other hand, in the case of the self-attractive nonlinearity (on the
average), we observe the opposite phenomenon: the application of the FRM
results in untrapped oscillations, if the unmanaged condensate was
self-trapped in one of the two wells. If the periodically modulated
nonlinearity does not change its sign, the frequency of the untrapped
oscillations supported by the FRM is almost exactly equal to a quarter of
the driving frequency, suggesting a higher-order parametric resonance as an
underlying mechanism. 
It is worth mentioning here that, this frequency relation, is
relevant, as in the previous case, for sufficiently low densities. 

Our results were corroborated with ones derived in the framework of a 
semi-analytical approach. The latter, was based on an a Galerkin-type expansion that was used 
to describe the evolution of the wavefunctions at each well of the DWP. This led to 
a non-autonomous system of ODEs, which were solved numerically to find 
the relative difference in the atomic population between the two wells. 
In all cases, the ODE result was compared to the one obtained by the GPE model, 
and they were found to be in fairly good agreement: in particular, 
we found a quantitatively good agreement for small values of the modulation 
depth, $g_1<1$ (in both repulsive and attractive cases),
and a qualitatively good agreement for $g_1>1$ (for the repulsive case). 
Our approximation was found to fail only for $g_1>1$ in which case
the strong nonlinearity causes the involvement of additional modes
in the dynamics.

Detailed analysis of the resonances that may explain the character of the
FRM-induced trapping and detrapping is a nontrivial problem, because the
%simplest 
two-mode approximation, such as the one presented above or the one 
recently proposed in Ref. \cite{xie}, is insufficient for this purpose. 
The development of a more sophisticated finite-mode approximation, that may help uncover the
resonances, is a subject for a separate study. Another challenging
possibility is to consider similar dynamical effects in multidimensional
settings. It may also be interesting to extend the analysis to the case of a
binary BEC mixture trapped in a DWP. These problems will also be considered
elsewhere.

\textbf{Acknowledgments.} B.A.M. appreciates a support from the German-Israel Foundation through grant No.
149/2006. P.G.K. gratefully acknowledges the support of NSF-DMS-0349023
(CAREER), NSF-DMS-0806762 and of the Alexander von Humboldt Foundation. 
The work of D.J.F. was partially supported by the Special Account for Research Grants of the University of Athens.

\end{document}